\documentclass{article}

\usepackage[utf8]{inputenc}
\usepackage[sorting=none, giveninits=true]{biblatex}
\usepackage{graphicx}
\usepackage{subcaption}
\usepackage{amsmath}
\usepackage{tikz}
\usetikzlibrary{arrows.meta}
\title{Modal Analysis of Buffet Effects Induced by Ultrahigh Bypass Ratio Nacelle Installation}
\author{Sebastian Spinner\footnote{DLR, German Aerospace Center, Institute of Aerodynamics and Flow Technologies, 38108 Braunschweig, Germany, sebastian.spinner@dlr.de (Corresponding Author)} \and Andre Weiner\footnote{Technical University of Dresden, Institute of Fluid Mechanics, 01062 Dresden, Germany}}
\date{May 2026}

\begin{document}

\maketitle

\begin{abstract}
    Unsteady shock–boundary-layer interaction on the lower surface of a transport-aircraft wing can be caused or amplified by ultrahigh-bypass-ratio underwing nacelle installation.
    This work analyzes the resulting buffet dynamics on the Airbus XRF1 configuration at a Mach number of $0.84$, a Reynolds number of $3.3\times 10^6$, and $-4^\circ$ angle of attack using scale-resolving delayed detached eddy simulations and unsteady pressure-sensitive paint measurements.
    Coherent structures are extracted employing a data-efficient multi-taper spectral POD.
    Dominant modes occur in the Strouhal number range $St \in [0.1,0.3]$.
    Surface modes reveal wave-like shock motions that originate near the pylon–wing intersection and propagate inboard toward the fuselage.
    These shock oscillations are linked to unsteady flow separation downstream of the shock.
    Additional dominant modes show spanwise oscillations of the separated flow region and shock oscillations phase-linked to shear layer instabilities.
    The modal analysis of volume data reveals pressure waves connected to these modes traveling upstream above and below the wing.
\end{abstract}


\section{Introduction}
Modern transport aircraft design has to maximize efficiency to mitigate its climate impact and be competitive in the global market. However, uncertainties associated with transonic shock buffet at the high-speed border of the flight envelope necessitate significant safety margins during aircraft design. These additional margins directly contribute to increased aircraft empty weight, leading to higher fuel consumption and reduced payload capacity. A detailed understanding of the buffet phenomenon will enable informed design recommendations for future aircraft, improving efficiency while maintaining high levels of safety.

Although many decades have passed since the first description of the buffet phenomenon by Hilton and Fowler \cite{Hilton1947a}, the field of buffet research is still very active. While these early studies approached the problem primarily by means of experiments on 2D airfoils, numerical investigations have become increasingly important over the years. Experimental investigations and wind tunnel tests continue to play an important role, as they provide indispensable validation data for simulations. The configurations investigated became more and more complex over time, including finite span unswept \cite{Benoit1987} and swept wing models \cite{Steimle2012,Iovnovich2015,Lusher2026} up to full wing-body configurations \cite{Roos1985, Dandois2013,Koike2016,Apetrei2020,Goc2025}, until today, highly accurate simulations and wind tunnel tests are carried out on entire aircraft models, including nacelles and tailplane \cite{Lutz2023}.

Buffet analysis is challenging due to the generation of broadband, three-dimensional signals that are difficult to decouple.
Consequently, isolating the underlying physical mechanisms driving the phenomenon is hard.
In the past decade, modal analysis techniques emerged as a promising approach to support the analysis and visualization of spatially and temporally resolved snapshot data.
The most commonly employed techniques are proper orthogonal decomposition (POD), discrete Fourier transformation (DFT), dynamic mode decomposition (DMD), spectral POD (SPOD), and resolvent analysis.
Taira et al. \cite{Taira2017} provide an introduction and overview.
A theoretical connection between SPOD, DMD, and resolvent analysis is established in \cite{Towne2018}.
While the POD is straightforward to compute and analyze, the modes may contain a mixture of phenomena associated with different frequencies or frequency bands, since POD modes maximize variance but not necessarily coherence \cite{Mendez2019}.
This spectral mixing can make the physical interpretation of modes more challenging.
The DMD builds on the POD by fitting a linear dynamics model within an approximately invariant subspace spanned by the leading POD modes.
The resulting decomposition is more interpretable since each spatial mode is associated with a unique frequency and growth rate.
However, in practice, DMD computations are very sensitive to imperfections in the data (noise, nonlinear dynamics, limited number of snapshots), leading to spuriously decaying or growing dynamics \cite{Weiner2023,Weiner2025}.
The SPOD according to \cite{Towne2018} combines DFT and POD.
In the vanilla SPOD algorithm, spectral estimates are obtained by computing the DFT of temporally overlapping blocks of snapshots.
A POD of the estimates for each frequency bin yields the sorted SPOD modes and their energies.
Applying the standard SPOD to simulation data is challenging due to the enormous volume of data that must be processed and stored.
Moreover, the sequence length and hence the DFT frequency resolution of simulation data are often very limited.
The frequency resolution is reduced further when splitting the sequence into blocks.
More recent SPOD variants with taper-based spectral estimates mitigate this issue and are computationally more feasible \cite{Schmidt2022,Yeung2024}.

The number of scientific studies employing modal analysis to buffet on full wing-body configurations is rather limited.
Ohmichi et al. \cite{Ohmichi2018} employ POD and DMD to study buffet on the NASA Common Research Model.
In the study, the authors analyze the 3D velocity field about the blank wing.
The snapshots are generated using a zonal detached eddy simulation (ZDES).
The dominant fluid structures oscillate in the range $St \in [0.2,0.6]$ ($St=fc_\mathrm{MAC}/U_\infty$ - Strouhal number, $f$ - frequency, $c_\mathrm{MAC}$ - mean aerodynamic chord, $U_\infty$ - free stream velocity) and propagate from mid span towards the wing tip.
The propagating shock oscillations create a cellular mode shape with increasing length towards the tip.
The authors also identify low-frequency disturbances with $St\approx 0.06$, which modify the entire shock front nearly uniformly, similar to 2D airfoil buffet.
Masini et al. \cite{Masini2019} studied the RCB12 half model with blank wing in buffet conditions by means of unsteady pressure sensitive paint (uPSP).
The authors identify two distinct phenomena dominating the flow physics, namely, low-frequency shock unsteadiness in the range $St \in [0.05,0.15]$, connected to mostly inboard traveling pressure disturbances, and broadband outboard running pressure waves in the range $St \in [0.2,0.5]$, forming the buffet cells.
In a follow-up study by the same authors \cite{Masini2020}, they supplement the wind tunnel tests with scale-resolving delayed detached eddy simulations (DDES).
Comparing the modal analyses of experimental and numerical data reveals that the simulations successfully reproduce the broadband, outboard-propagating pressure disturbances, centered at $St \approx 0.27$.
However, the low-frequency content remains unique to the experimental data.
Among others, the authors hypothesize the wing's flexibility and wind tunnel disturbances as potential discrepancies between experiments and simulations.
Ohmichi and Sugioka \cite{Ohmichi2026} apply variational mode decomposition-based nonstationary coherent structure (VMD-NCS) analysis to unsteady pressure-sensitive paint measurements on the NASA common research model. The method isolates low-frequency shock oscillations ($St \approx 0.1$) and buffet cells ($St \in [0.37, 0.8]$). Their analysis reveals that buffet cell amplitude is not directly coupled to the phase of shock oscillations. Instead, buffet cell fluctuations correlate with the boundary layer separation state, persisting strongly in regions of sustained separation and varying with intermittent separation outboard. These experimental findings support computational studies suggesting buffet cells emerge from separation mechanisms rather than direct shock feedback \cite{Plante2021}.

While sophisticated experimental and numerical techniques allow the investigation of increasingly complex configurations, many questions concerning buffet and its potential interplay with other phenomena remain open.
Therefore, the FOR 2895 research project was initiated to improve the understanding of buffet phenomena on realistic transport aircraft configurations. The initiative involves detailed numerical and experimental investigations of unsteady aerodynamic phenomena occurring at the border of the flight envelope \cite{Lutz2023}.
A special research area addressed in the project is unsteady shock boundary layer interaction on the wing's lower surface due to ultra high bypass ratio (UHBR) engine installation. The trend towards large nacelles requires them to be coupled closely to the wing to mitigate adverse engine installation effects. For under-wing mounted engines, this requirement results in the formation of a half-open channel between nacelle, pylon, wing lower surface, and fuselage, leading to additional flow acceleration and, under certain conditions, to shock buffet phenomena. 

The UHBR-induced shock buffet phenomenon was first analyzed in detail in a previous study based on uPSP measurements obtained during wind tunnel testing in the European Transonic Windtunnel (ETW) \cite{Spinner2024}.
It was found that unsteady shock wave-boundary-layer interaction occurs at low angles of attack and high subsonic Mach numbers.
The analysis showed that the phenomenon is present over the entire investigated range of Reynolds numbers $3.3\times 10^6 \leq Re \leq 25\times 10^6$ ($Re_\infty = U_\infty c_\mathrm{MAC}/\nu_\infty$, $\nu_\infty$ - free stream kinematic viscosity).
Spectral analysis indicated that the UHBR-induced buffet behaves similarly to the buffet on the wing upper surface at high angles of attack.
A follow-up study investigated the UHBR-induced shock buffet using high-fidelity improved delayed detached eddy simulations (IDDES) \cite{Spinner2025}.
The IDDES accurately reproduced the surface pressure mean and variance of the wind tunnel experiment. The power spectral density (PSD) of a local time series of pressure data extracted at the shock location showed good agreement between the IDDES data and the previous study based on uPSP measurements, providing another indication that the lower wing buffet phenomenon due to UHBR engine installation may be linked to the Strouhal number range of $St \in [0.2,0.4]$.

Having identified the Strouhal-number range relevant to lower wing buffet in both experimental and numerical investigations, the question arises as to which aerodynamic structures occur. To answer this question,  this study expands the comparison of experimental and numerical data through modal analysis. For the numerical data set, DDES was employed due to its favorable computational efficiency, which facilitated the generation of extended time series and improved frequency resolution. Moreover, the numerical data allow performing a modal analysis of the flow field in the volume, something that cannot be done with the experimental pressure sensitive paint dataset.
The test case and simulation setup are summarized in Section \ref{sec:test_case}, while Section \ref{sec:spod} outlines the modal analysis techniques applied to maximize coherence within the extracted modes. The results of this analysis are then presented in Section \ref{sec:results} and discussed in Section \ref{sec:discussion}, providing a detailed evaluation of the dominant flow phenomena.

\section{Test Case}
\label{sec:test_case}
\begin{figure}
  \centering
  \begin{minipage}{0.45\textwidth}
    \centering
    \includegraphics[width=\linewidth]{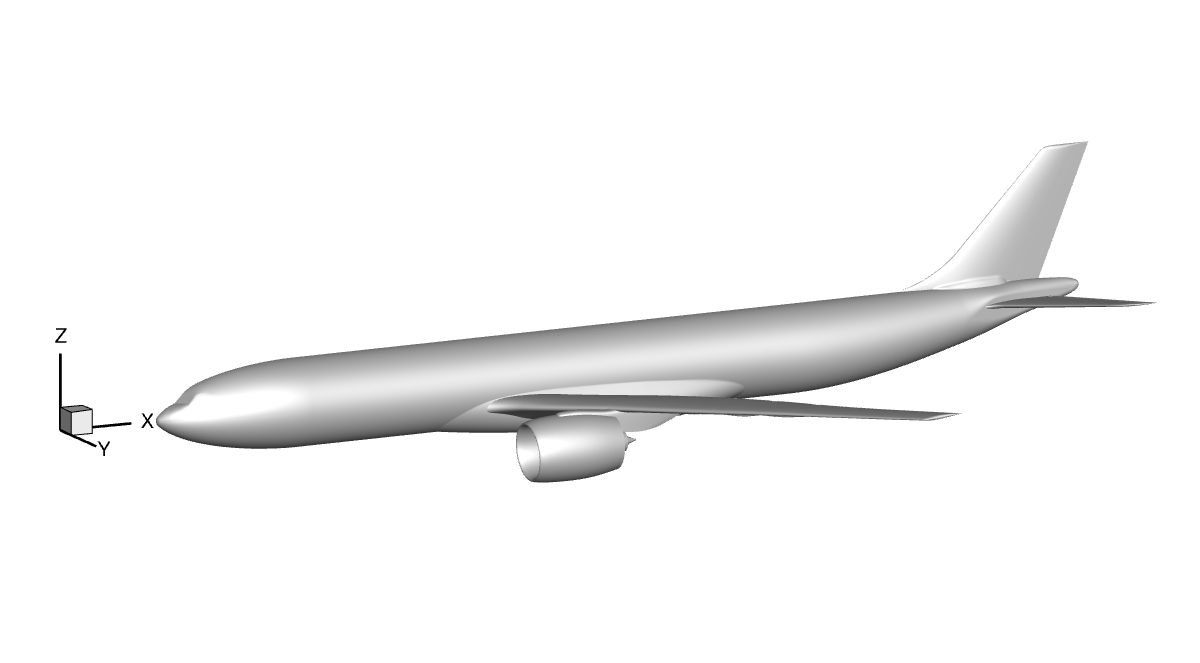}
    \subcaption{CAD geometry of half model used in numerical simulations}
    \label{fig:xrf1_cad}
  \end{minipage}
  \hspace{0.05\textwidth}
  \begin{minipage}{0.45\textwidth}
    \centering
    \includegraphics[width=\linewidth,trim={0 8cm 0 12cm},clip=True]{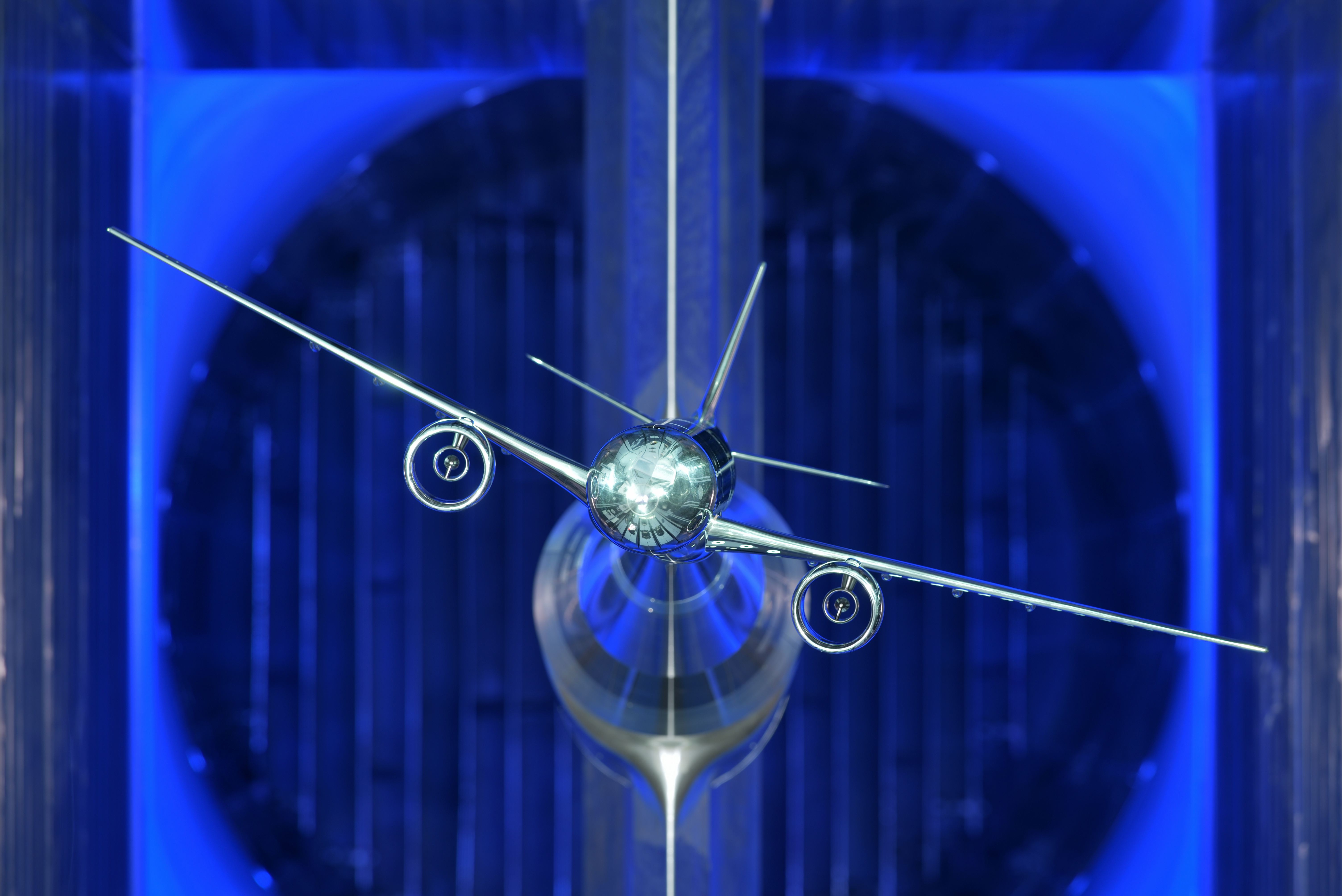}
    \subcaption{Wind tunnel model in ETW test section}
    \label{fig:xrf1_etw}
  \end{minipage}
  \caption{Airbus XRF1 research aircraft}
  \label{fig:setup_overview}
\end{figure}

The geometry investigated is the Airbus XRF1 wind tunnel model \cite{Mann2019} equipped with UHBR through flow nacelles \cite{Spinner2021}, as shown in Fig. \ref{fig:setup_overview}. It is representative of a state-of-the -art long-range transport aircraft and was designed for high speed testing in cryogenic wind tunnel facilities. The reference flow conditions of this study are motivated by the available wind tunnel test points as well as the high computational demands for scale-resolving simulations at high Reynolds numbers. The data analyzed in this study was measured and simulated at a Reynolds number of $3.3 \times 10^6$, a Mach number of $0.84$, and an angle of attack of $-4$°. The selected angle of attack in this study is unexpectedly low, constraining the flight envelope at high cruise Mach number and aligning with FAA design speed specifications. Such conditions are critical for demonstrating $V_{DF}$ (demonstrated flight diving speed) \cite{Part25-335}, making understanding of associated aerodynamic behavior and buffet onset essential.

\subsection{DDES}
\begin{figure}
  \centering
  \begin{minipage}{0.45\textwidth}
    \centering
    \includegraphics[width=\linewidth]{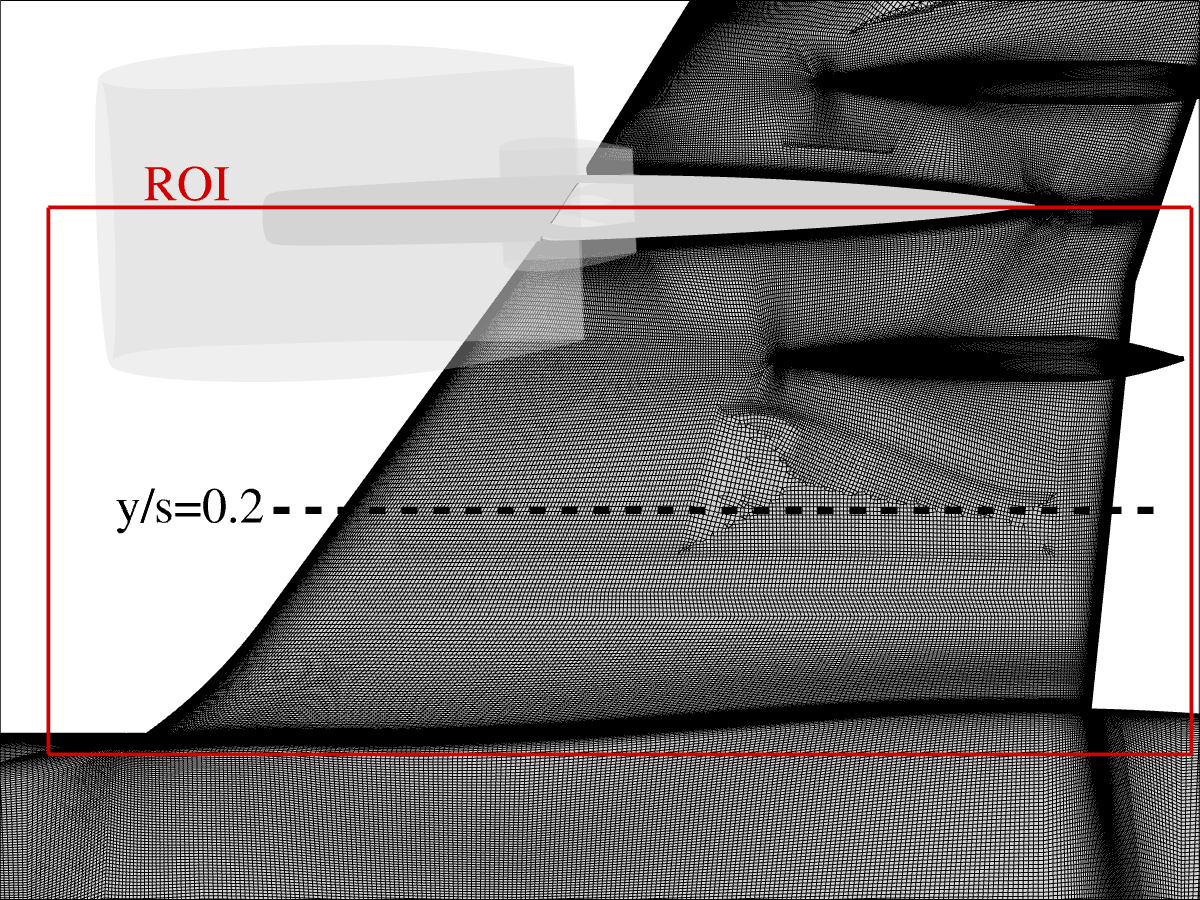}
    \subcaption{Surface mesh in region of interest (ROI)}
    \label{fig:surf_mesh}
  \end{minipage}
  \hspace{0.05\textwidth}
  \begin{minipage}{0.45\textwidth}
    \centering
    \includegraphics[width=\linewidth]{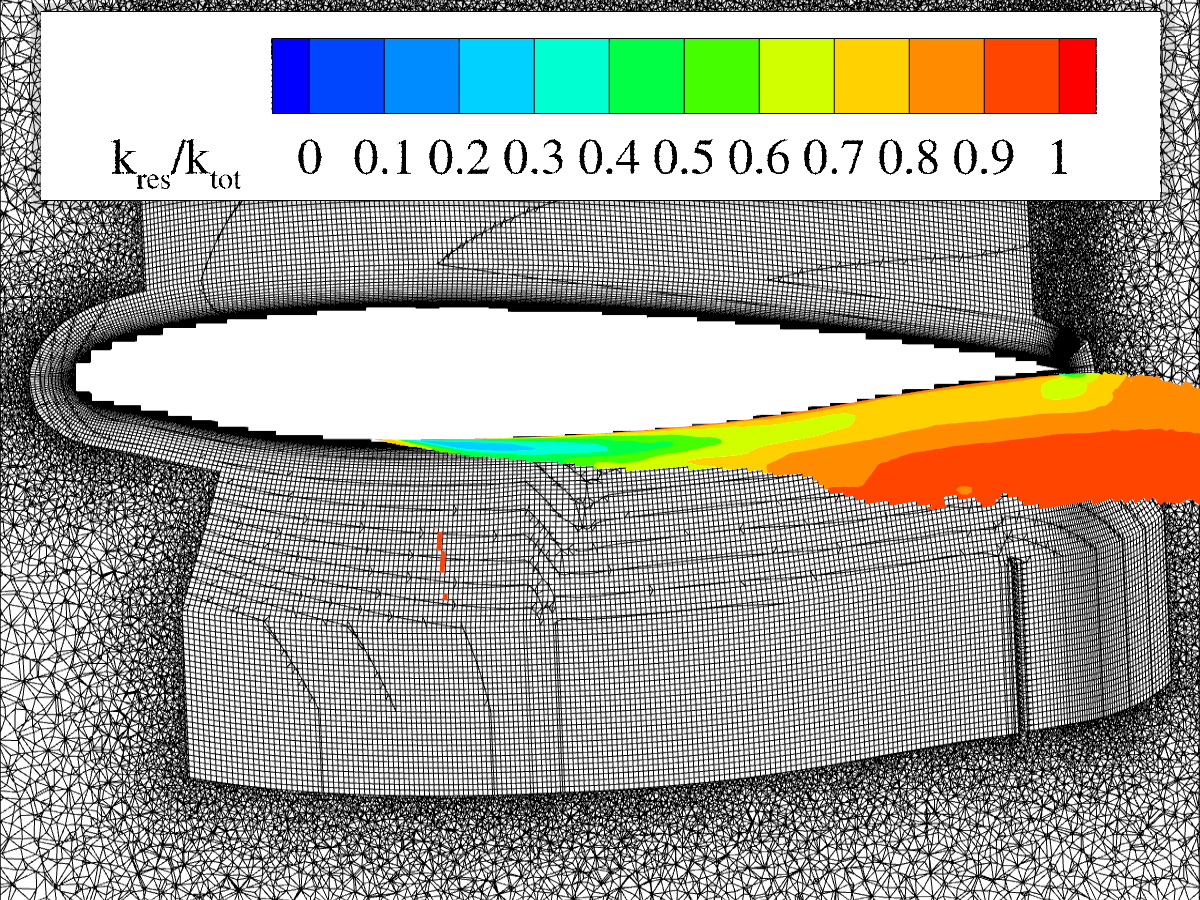}
    \subcaption{Volume mesh at $y/s\approx0.2$}
    \label{fig:vol_mesh}
  \end{minipage}
  \caption{Details of the DDES mesh (airfoil geometry modified, non-representative of XRF1)}
  \label{fig:ddes_mesh}
\end{figure}

The simulation setup was similar to our previous IDDES study \cite{Spinner2025}, assuming symmetric flow and running the simulation for a half model of the aircraft. However, due to the less stringent mesh resolution requirements of DDES, the simulation was run on a separate mesh with 112 million points \cite{Lutz2023}, which was also used as the basis for refinement in the IDDES simulation. Details on the mesh are shown in Fig. \ref{fig:ddes_mesh} (the airfoil geometry has been concealed for reasons of confidentiality). The surface mesh on the wing lower surface as depicted in Fig. \ref{fig:surf_mesh} was created from predominantly isotropic quadrilaterals with an maximum spacing of $h_{\mathrm{max}} < 0.01 \, c_{\mathrm{MAC}}$ and an average spacing of $h_{\mathrm{ave}}<0.007 \, c_{\mathrm{MAC}}$. The first layer height was set to ensure $y^{+}<0.4$, and a boundary layer mesh was grown with a wall normal growth factor of 1.12 up to a wall normal height of two times the estimated maximum boundary layer thickness based on flat plate theory. To enhance the resolution of transonic shocks, additional mesh blocks of mostly isotropic hexahedra with an average edge length of $h_{\mathrm{ave}}<0.01 \, c_{\mathrm{MAC}}$ were created at the edge of the boundary layer mesh above and below the wing as shown in Fig. \ref{fig:vol_mesh} for a slice through the mesh at $y/s=0.2$ ($y$ - y-coordinate, $s$ - model half span). To support that this mesh resolution is sufficient, Fig. \ref{fig:vol_mesh} additionally shows contours of the ratio of resolved ($k_{res}$ - resolved turbulent kinetic energy) to total turbulent kinetic energy ($k_{tot}$ - total turbulent kinetic energy) for regions where $Tu_{loc}>0.1$ ($Tu_{loc}=\sqrt{\frac{2}{3}k_{res}} / V_{abs}$ - local resolved turbulence intensity, $V_{abs}$ - absolute Velocity), as suggested by \cite{Probst2025}. The contour of $\frac{k_{res}}{k_{tot}}$ shows that downstream of the shock (cf. Fig. \ref{fig:ddes_mach_slice}), resolved turbulence starts to develop and the resolved portion of turbulence grows quickly until further downstream the majority of turbulent fluctuations are resolved.

As in our IDDES study \cite{Spinner2025} the wing geometry was deformed by applying twist and bend distributions measured in wind tunnel tests. To achieve this, mesh deformation was employed to accurately match the shape of the physical model tested in the wind tunnel under corresponding aerodynamic loads.

The computational domain was set up as a half model with a symmetry plane at $y=0$ and spanned
\[
\frac{x}{c_{\mathrm{MAC}}} \in \left[-407,\;407\right], \qquad
\frac{y}{c_{\mathrm{MAC}}} \in \left[0,\;407\right], \qquad
\frac{z}{c_{\mathrm{MAC}}} \in \left[-407,\;407\right]
\]

with cartesian coordinates $x$, $y$ and $z$ along the aircraft longitudinal, spanwise and upright direction (cf. Fig. \ref{fig:xrf1_cad}).
A farfield boundary condition was set at the outer boundary of the domain and no-slip condition was applied to the walls of the aircraft model. The simulation was run fully turbulent without any laminar-turbulent transition effects. 

A hybrid RANS-LES method based on a Reynolds-stress RANS model \cite{Herr2023IDDESRSM} is utilized, which in the present work is operated in a DDES mode \cite{Probst2011}.
The subdivision of the flow field into RANS and LES regions was based purely on the algorithm of the DDES, no geometrical constraints for RANS and LES regions were enforced. The simulation was conducted using the unstructured cell-vertex based flow solver TAU \cite{GerholdTAU}, with a second-order central scheme and matrix dissipation employed for the mean flow equations, and a first-order Roe upwind scheme used for the turbulence equations. To improve accuracy in the DES regions while maintaining stability in the RANS regions, a hybrid low-dissipation low-dispersion scheme \cite{Probst2016} was applied. The SSG/LRR Reynolds-stress model \cite{Eisfeld2016} with an ln-\ensuremath{\omega} formulation for the length scale equation was used in the RANS region, along with a simple gradient diffusion model \cite{Eisfeld2004} and an isotropic dissipation model. The simulation utilized a dual time stepping scheme with a backward Euler relaxation solver, and a time resolution of 1000 time steps per CTU ($CTU=c_\mathrm{MAC}/U_\infty$ - convective time unit) was selected. The latter resulted in convective CFL numbers below unity in most cells in the DES region, which is deemed a prerequisite for accurate LES simulations \cite{Spalart2001}. Cauchy convergence criteria were applied to ensure convergence of the inner iterations of the dual time stepping scheme and accelerate the simulation. Surface pressure data was acquired at a sampling frequency of 13.73 kHz, corresponding to a Strouhal number of 10.0, and a total of 500 surface snapshots were recorded over a time interval of 36.4\, ms (50 CTU). Over the same time interval, 100 volume snapshots were recorded with a sampling frequency of 2.674 kHz, corresponding to a Strouhal number of 2.0. These sampling frequencies are considered sufficient because previous studies \cite{Spinner2024,Spinner2025} suggest that the buffet phenomenon occurs in the Strouhal‑number range 0.2 to 0.4. Accordingly, the sampling frequency of the surface data is at least 20 times higher, and that of the volume data is at least 2 times higher than the highest expected dominant frequency.
Since the DDES exhibited transient phenomena at the start of the simulation, the data from the first 5 CTU were excluded from all subsequent analyses.

\subsection{Wind tunnel testing}
Experimental data was obtained from cryogenic wind tunnel tests conducted in the ETW. 
The lower wing of the model was coated with a pressure sensitive paint with a thickness of no more than 5$\mu$m. Steady measurements were captured using the lifetime method with a gated CCD camera and pulsed UV LED lights \cite{Yorita2018}. Unsteady measurements employed the intensity method via a high-speed CMOS camera and constant-power LEDs \cite{Klein2020}. The reconstruction of unsteady absolute pressures was achieved in post-processing by superposition of the steady and the unsteady measurements \cite{Yorita2024}, resulting in time-resolved pressure information on the wing surface. The sampling frequency of the uPSP measurement was 2 kHz, corresponding to a Strouhal number of 1.456. In total, 3600 snapshots over a time interval of 1.8s (2472 CTU) were recorded, of which the first 3500 snapshots were used in the analysis. While a comprehensive uncertainty quantification for the uPSP measurements is challenging due to the complexity of the cryogenic environment, the reliability of the data is demonstrated by the excellent agreement with static pressure tappings (uncertainty lower than 0.5\%) and unsteady Kulite sensors (uncertainty lower than 0.1\%), and by verifying the physical consistency of the observed buffet trends.
For further details on the conducted wind tunnel tests, the reader is referred to \cite{Spinner2024, Lutz2023, Waldmann2023}.

\section{Spectral proper orthogonal decomposition}
\label{sec:spod}

In this work, we employ the recently proposed adaptive taper-based SPOD by Yeung and Schmidt \cite{Yeung2024}.
The algorithm and practical aspects are briefly summarized in this section.
The implementation is available as open source software \cite{Weiner2021}.

Suppose the normalized pressure fluctuations of the $n$th snapshot in the region of interest are arranged in a state vector $\mathbf{x}_n$ of length $M$.
The number of discrete spatial locations $M$ and the number of available snapshots $N$ vary across the different datasets, but in all cases $M\gg N$.
The sequence of flow states is then arranged into a data matrix of shape $M\times N$:
\begin{equation}
    \label{eq:data_matrix}
    \mathbf{X} = \mathrm{diag}(\sqrt{\mathbf{w}})\left[\mathbf{x}_1, \mathbf{x}_2,\ldots , \mathbf{x}_N\right].
\end{equation}
The weighting vector $\mathbf{w}$ is motivated by the definition of the inner product and, in practice, reduces the mesh-induced bias of simulation data \cite{Weiner2023}.
The definition of $\mathbf{w}$ depends on the dataset:
\begin{itemize}
    \item DDES surface data: $\mathbf{w}$ is formed using the area of surface elements
    \item DDES volume data: $\mathbf{w}$ is formed using the cell volumes
    \item uPSP data: $\mathbf{w}$ is a binary mask that sets spurious pixel values to zero
\end{itemize}
For tall-and-skinny matrices, it is computationally more efficient to perform the SPOD in the POD subspace, i.e., to first project the data onto its POD basis:
\begin{equation}
    \label{eq:subspace_data}
    \tilde{\mathbf{X}} = \mathbf{U}^T\mathbf{X} = \mathbf{\Sigma V}^T,\quad
    \mathbf{X} = \mathbf{U\Sigma V}^T,
\end{equation}
where $\mathbf{U}$, $\mathbf{\Sigma}$, and $\mathbf{V}$ are the left-singular vectors, singular values, and right-singular vectors of the economy singular value decomposition, respectively.
Note that analyzing the projected data matrix of size $N\times N$ is equivalent to analyzing the full $M\times N$ data \cite{Yeung2024}, i.e., the resulting spectra are identical and the full-state modes are recovered by projection onto the POD basis (details follow below).

Since the sequence lengths of all datasets are limited, we employ a purely taper-based spectral estimation.
The Fourier-transformed state vector $\hat{\mathbf{x}}_k^{(n)}$ of the $k$th frequency bin employing the $(n)$th taper function is given by \cite{Yeung2024}:
\begin{equation}
    \label{eq:tapered_dft}
    \hat{\mathbf{x}}^{(n)}_{k} = \sum_{j=0}^{N_{\mathrm{fft}}-1} v^{(n)}_{j+1}\,\tilde{\mathbf{x}}_{j+1}\, e^{-\mathrm{i} 2\pi j k / N_{\mathrm{fft}}},
\end{equation}
with $k = 0,\ldots,N_{\mathrm{fft}}-1$ and $(n) = 1,\ldots,N_{\mathrm{win}}$.
We use $N_\mathrm{fft}=3500$ for all datasets, which equals the number of uPSP snapshots used in the analysis.
Consequently, the shorter numerical datasets are zero-padded.
We found that keeping $N_\mathrm{fft}$ constant simplified visual comparisons between spectra.
The number of tapers $N_{\mathrm{win}}$ is comparable to the number of blocks in the vanilla SPOD and will be discussed further at the end of this section.
The taper weight $v^{(n)}_{j}$ reads \cite{Yeung2024}:
\begin{equation}
    \label{eq:taper_weights}
    v^{(n)}_{j}	= \sqrt{\frac{2}{N_{\mathrm{fft}} + 1}}\, \sin\!\left(\frac{\pi n j}{N_{\mathrm{fft}} + 1}\right).
\end{equation}
Employing sinusoidal tapers yields another computational advantage since the weighted Fourier transforms can be computed with ease given a single unweighted DFT \cite{Yeung2024}.
The scaling factor in (\ref{eq:taper_weights}) preceding the sine-function ensures that the sum over all frequency bins yields one, i.e., the power is not shifted up or down due to the weighting. 

To combine the multi-taper estimates, the Fourier-transformed states of the $k$th frequency are arranged into an $N\times N_\mathrm{win}$ matrix:
\begin{equation}
    \label{eq:dft_data_matrix}
    \hat{\mathbf{X}}_k = \sqrt{\Delta t} \left[\hat{\mathbf{x}}^{(1)}_k, \hat{\mathbf{x}}^{(2)}_k, \ldots, \hat{\mathbf{x}}^{(N_\mathrm{win})}_k  \right]\mathrm{diag}\left(\left[\mu_1, \mu_2, \ldots, \mu_{N_\mathrm{win}}\right]^{1/2}\right),
\end{equation}
where the parabolic weights $\mu_{(n)}$ are computed as \cite{Yeung2024}:
\begin{equation}
    \label{eq:parabolic_weights}
    \mu_{(n)} = \frac{6}{4N_\mathrm{win}^3 + 3N_\mathrm{win}^2 - N_\mathrm{win}}
    \left(N_\mathrm{win}^2-(n-1)^2\right).
\end{equation}
By construction, the parabolic weights sum up to one.
The weights penalize higher-order tapers with larger bandwidths \cite{Yeung2024}.
The eigendecomposition of the cross-spectral density matrix yields the SPOD modes in the reduced space and the modal energies:
\begin{equation}
    \label{eq:spod_modes}
    \hat{\mathbf{X}}_k\hat{\mathbf{X}}_k^\ast \tilde{\mathbf{\Phi}}_k = \tilde{\mathbf{\Phi}}_k \tilde{\mathbf{\Lambda}}_k.
\end{equation}
In the case that $N > N_\mathrm{win}$, it is more efficient and numerically stable to solve the eigenvalue problem:
\begin{equation}
    \label{eq:eigen}
    \hat{\mathbf{X}}_k^\ast \hat{\mathbf{X}}_k \mathbf{\Psi}_k = \mathbf{\Psi}_k \mathbf{\Lambda}_k,
\end{equation}
where $\mathbf{\Lambda}_k$ is the non-zero part of $\tilde{\mathbf{\Lambda}}_k$ and the SPOD modes in the reduced space are given by $\tilde{\mathbf{\Phi}}_k = \hat{\mathbf{X}}_k \mathbf{\Psi}_k \mathbf{\Lambda}^{-1/2}_k$.
Finally, the SPOD modes in full state space are recovered by projecting the reduced modes onto the POD basis and undoing the initial weighting, i.e., $\mathbf{\Phi}_k = \mathrm{diag}(1/\sqrt{\mathbf{w}})\mathbf{U}\tilde{\mathbf{\Phi}}_k$.
Note that the weighting is not undone for the uPSP data to avoid division by zero.

Another advantage of multi-taper estimates is the possibility to select $N_\mathrm{win}$ adaptively per frequency.
Increasing the number of taper estimates reduces variance but increases bias.
Employing a constant $N_\mathrm{win}$ may lead to unnecessary bias in some parts of the spectrum, while others are not sufficiently converged.
As a remedy, Yeung and Schmidt \cite{Yeung2024} suggest selecting $N_\mathrm{win}$ based on the change of the leading SPOD mode:
\begin{equation}
    \label{eq:adaptive_taper}
    \alpha^{(n)}_{k} = \left| \left( \tilde{\mathbf{\phi}}^{(n)}_{1,k} \right)^\ast \tilde{\mathbf{\phi}}^{(n-1)}_{1,k} \right|,
\end{equation}
with the breaking criterion $1 - \alpha^{(n)}_{k} \leq tol$.
As an additional constraint, the difference in $N_\mathrm{win}$ between two neighboring bins is limited to one.
We set $tol=10^{-5}$ and limit the maximum number of tapers to $200$ and $400$ for the DDES and uPSP data, respectively.
Parameter studies conducted with regard to tolerance and maximum number of tapers showed that this parameter combination is best suited for isolating the dominant modes.
Limiting the maximum number of tapers avoids excessively long computations caused by individual frequency bins, typically at the upper end of the spectrum.
Overall, the spectra determined using SPOD proved to be very stable, and even with parameters deviating from the optimum, identical or very similar modes could be found.
Computing the adaptive taper-based SPOD took around 1 minute for the uPSP and about 3 minutes for the DDES data on an NVIDIA RTX 4000 workstation GPU.

\section{Results}
\label{sec:results}
The analysis of surface snapshots is limited to the area of interest on the inboard wing lower surface between pylon and fuselage. The data outside of the region of interest is blanked such that the data matrices for the modal analysis only include pressure data as depicted in Fig. \ref{fig:mean_cp}. Further preprocessing of the snapshots data included applying inter-quantile-range (IQR)  outlier detection to the uPSP data before computing the modal analysis. The IQR algorithm is applied per pixel with default parameters, and less than $0.2\%$ of all pressure values (space and time) are replaced based on the median of the six nearest neighbors (in time).

\subsection{Validation of DDES against Experimental Data}
\label{sec:ddes_validation}
\begin{figure}
    \centering
    \begin{subfigure}{0.49\textwidth}
        \begin{tikzpicture}
            \node[anchor=south west, inner sep=0] (image) at (0,0) {\includegraphics[width=\linewidth]{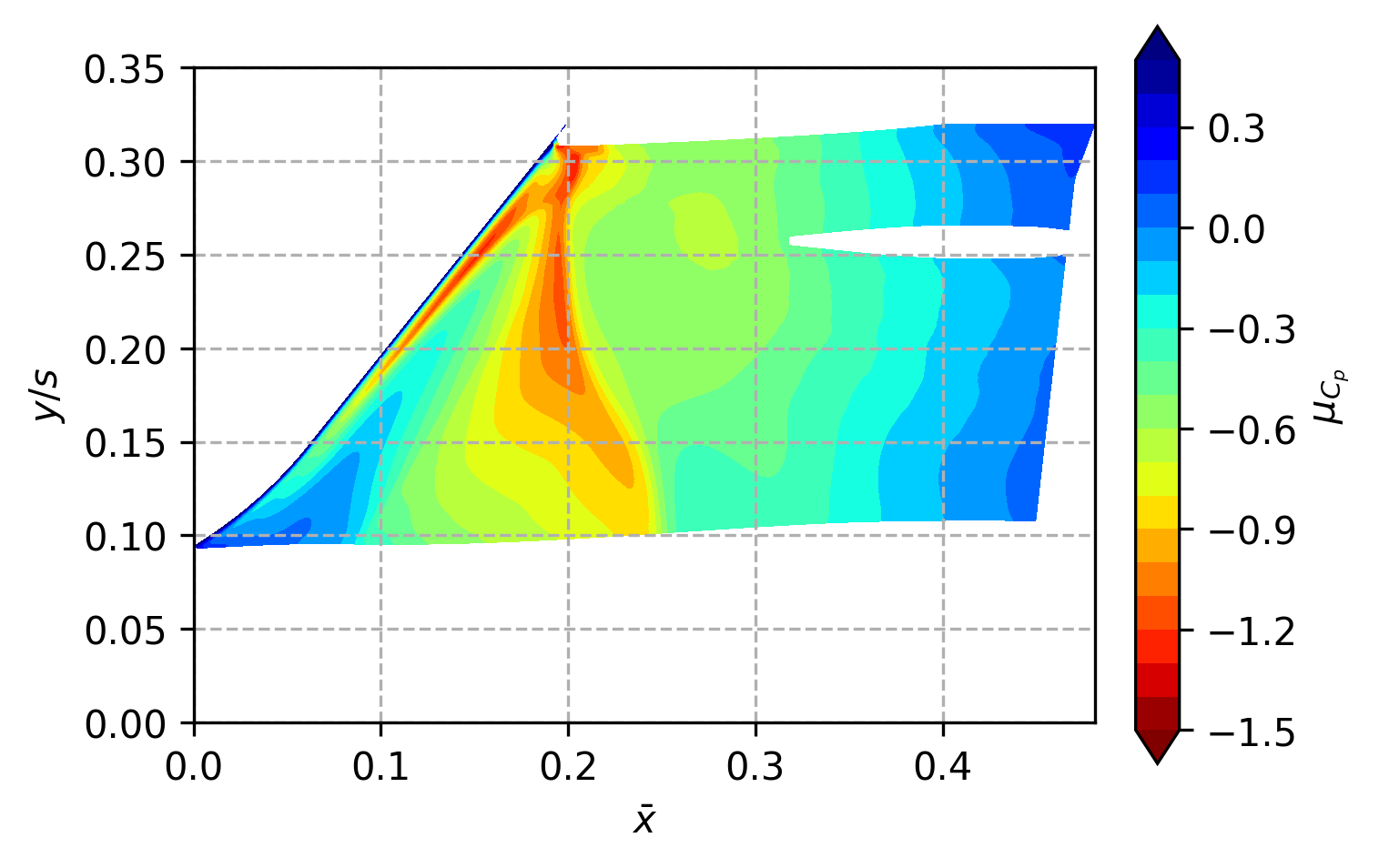}};
            \begin{scope}[x={(image.south east)}, y={(image.north west)}]
                \draw[->, line width=1pt, draw=black] (0.2, 0.7) -- (0.3, 0.7) node[midway, above, text=black] {Flow};
            \end{scope}
        \end{tikzpicture}
        \caption{DDES}
    \end{subfigure}
    \begin{subfigure}{0.49\textwidth}
        \includegraphics[width=\linewidth]{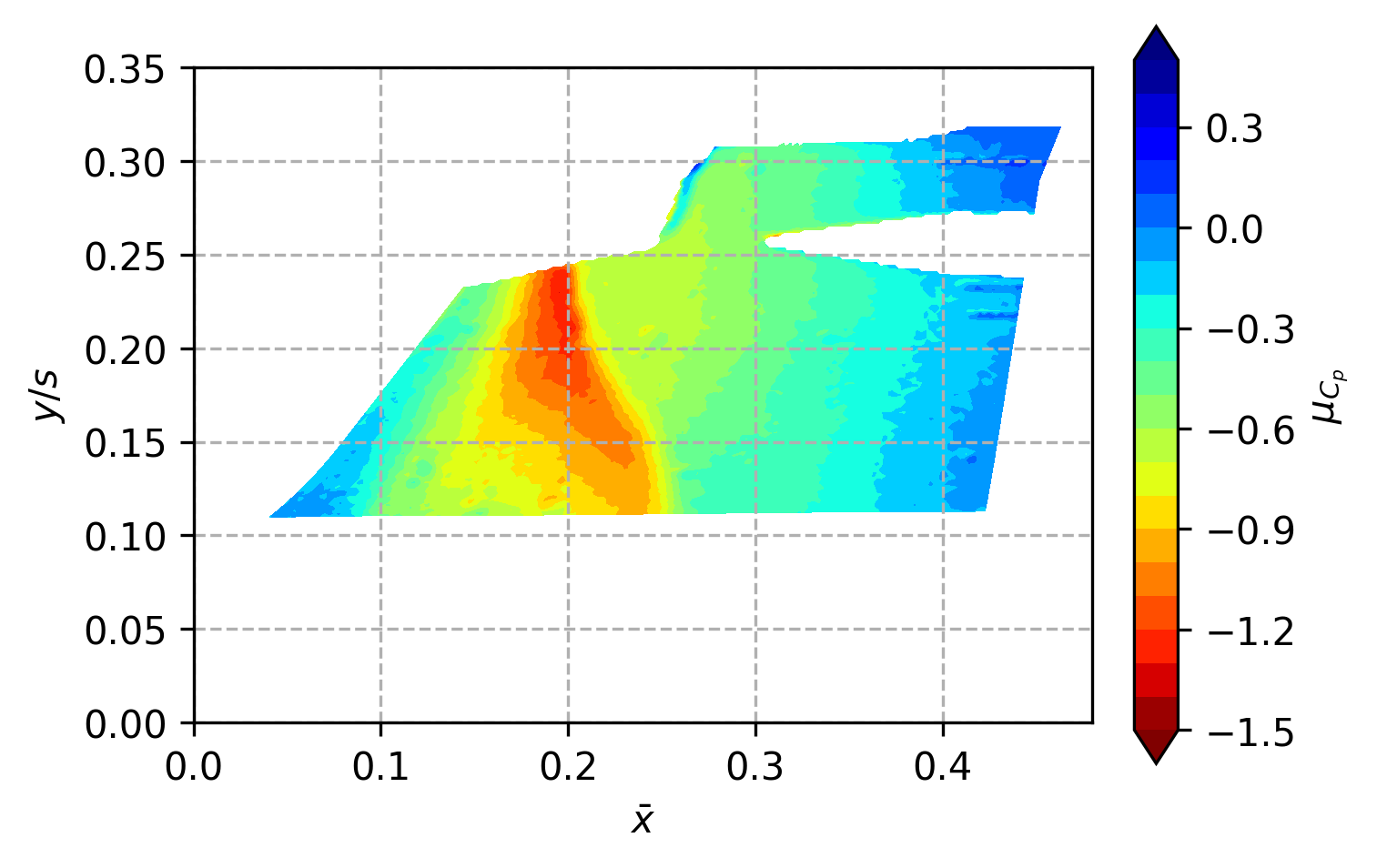}
        \caption{uPSP}
    \end{subfigure}
    \caption{Temporal mean of $C_p$ on the inboard wing lower surface}
    \label{fig:mean_cp}
\end{figure}

Figure \ref{fig:mean_cp} shows the temporal mean $\mu_{C_p}$ of the pressure coefficient $C_p$ for the DDES computation and the uPSP measurement on the inboard wing lower surface. In addition to the conventional nondimensional spanwise coordinate $y/s$, we also introduced a corresponding nondimensionalization of the streamwise x‑coordinate $\bar{x}=(x-x_{min})/(y_{max}-y_{min})$. Note that the uPSP plot misses data in the forward part of the airfoil for $0.25<y/s<0.31$. In this area, the line of sight of the uPSP camera, which recorded the model from below, was obscured by the nacelle. In both the simulation and the uPSP data sets, the region of the inboard flap track fairing at $y/s \approx 0.28$ is blanked as well because no reliable uPSP data could be obtained there during wind tunnel testing. This way, both data sets share a common setup concerning the respective aerodynamic surfaces, facilitating comparison during modal analysis. The agreement of the mean pressure coefficients between DDES and uPSP is excellent, exhibiting a strong shock extending from the fuselage up to the pylon.

\begin{figure}
    \centering
    \begin{subfigure}{0.49\textwidth}
        \begin{tikzpicture}
            \node[anchor=south west, inner sep=0] (image) at (0,0) {\includegraphics[width=\linewidth]{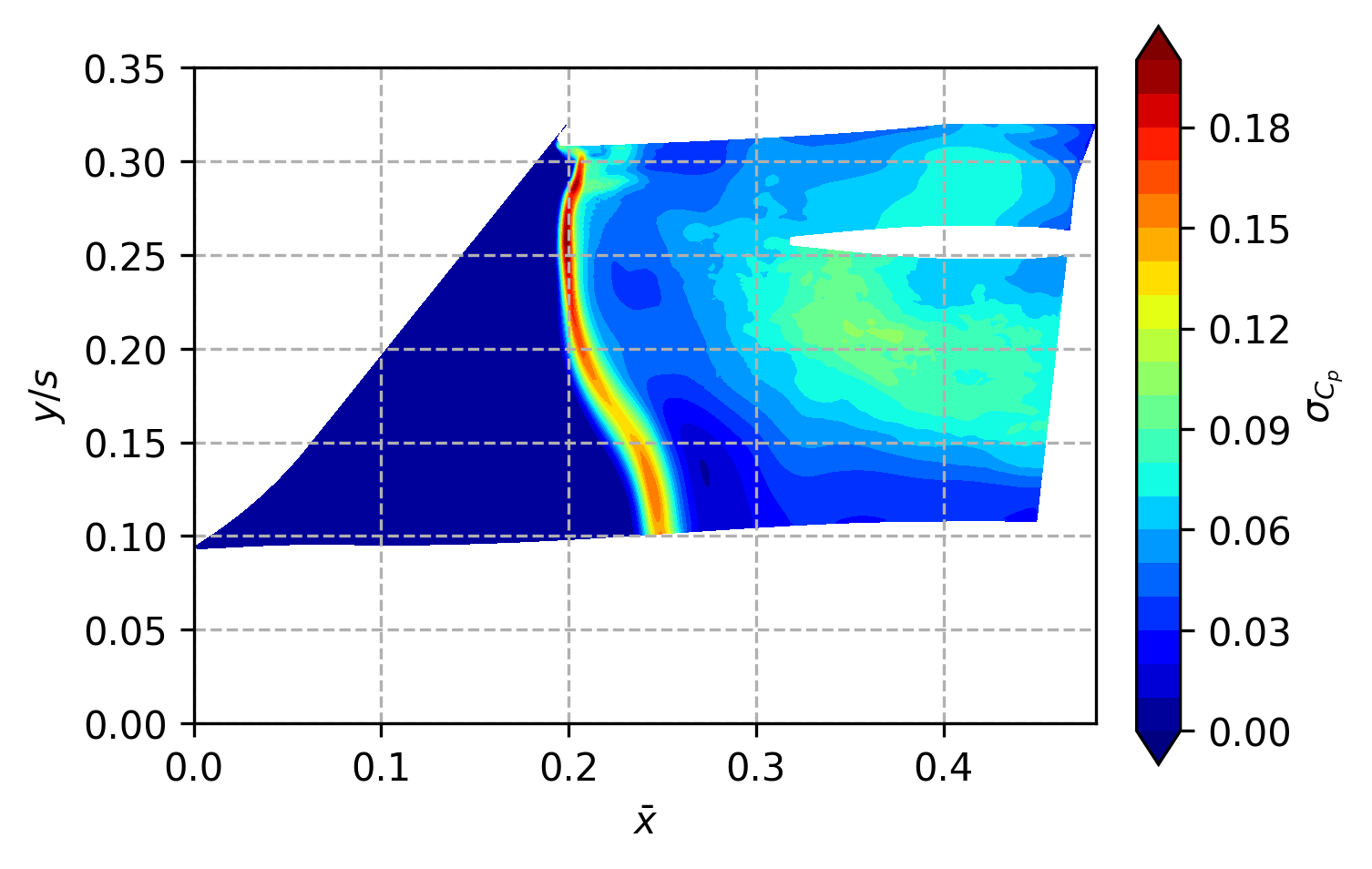}};
            \begin{scope}[x={(image.south east)}, y={(image.north west)}]
                \draw[->, line width=1pt, draw=black] (0.2, 0.7) -- (0.3, 0.7) node[midway, above, text=black] {Flow};
            \end{scope}
        \end{tikzpicture}
        \caption{DDES}
    \end{subfigure}
    \begin{subfigure}{0.49\textwidth}
        \includegraphics[width=\linewidth]{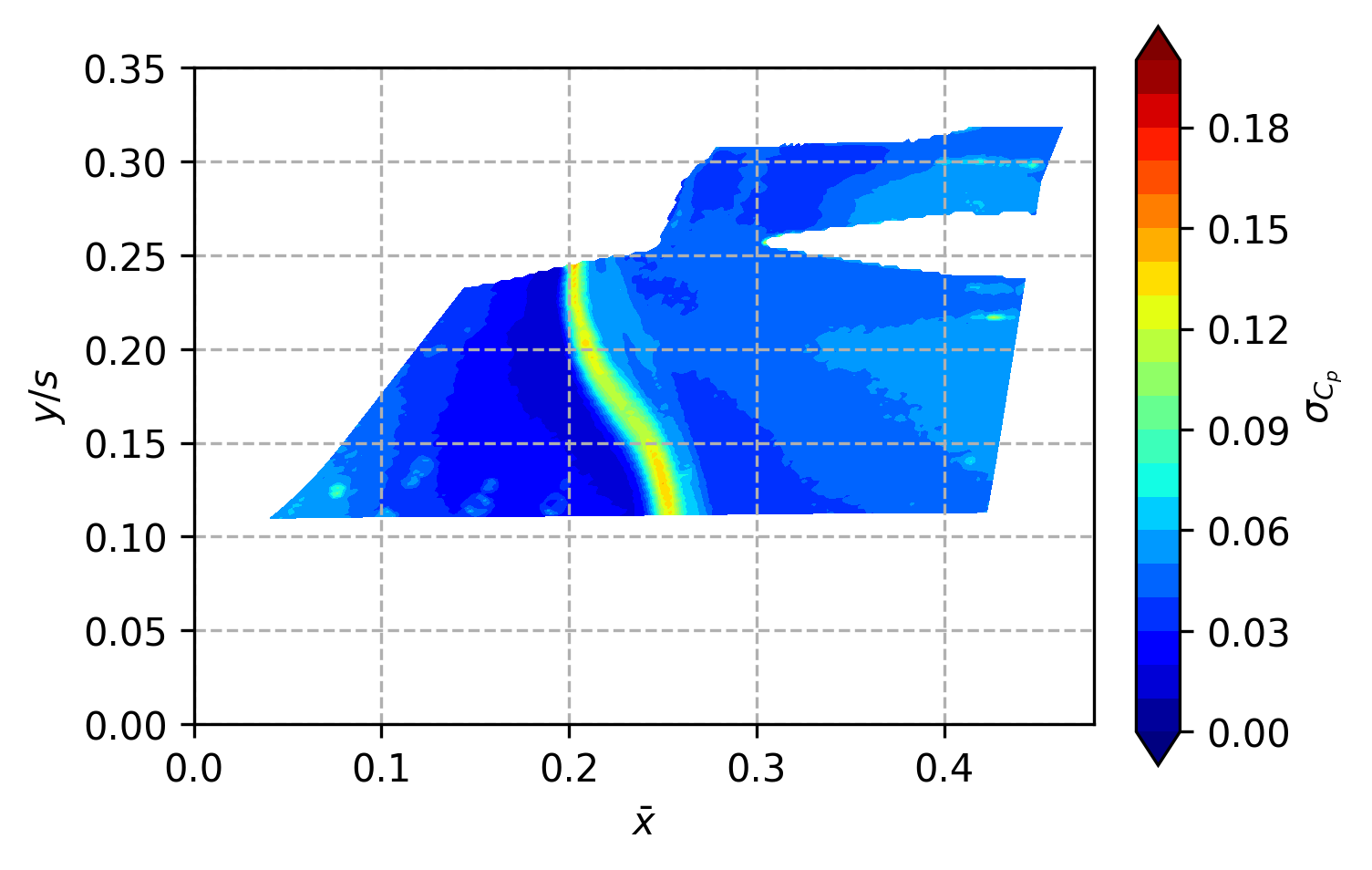}
        \caption{uPSP}
    \end{subfigure}
    \caption{Temporal standard deviation of $C_p$ on the inboard wing lower surface}
    \label{fig:var_cp}
\end{figure}

The unsteady nature of this shock front and the associated shock-induced separation can be assessed in Fig. \ref{fig:var_cp}, showing the temporal standard deviation of the pressure coefficient $\sigma_{C_p}$. Both the simulation as well as the experiment show high values of pressure fluctuation across the shock, indicating that the shock position is unsteady. Downstream of the shock, elevated values of $\sigma_{C_p}$ can be observed due to the occurring flow separation. It should be noted that these increased values are significantly more pronounced in the DDES data compared to the uPSP plot. The difference might arise due to the limited response time of the PSP paint itself or the limited sampling frequency of the uPSP measurements.
Although not shown explicitly here, the spatial surface pressure distribution obtained with the DDES simulations was found to agree very well with the previous IDDES simulation \cite{Spinner2025}, both in terms of temporal mean and standard deviation.

\begin{figure}
    \centering
    \begin{subfigure}{0.475\textwidth}
        \includegraphics[width=\linewidth]{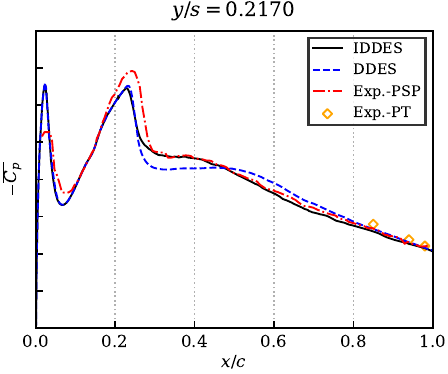}
        \caption{Temporal mean of $C_p$}
        \label{fig:slice_cp_mean}
    \end{subfigure}
    \begin{subfigure}{0.505\textwidth}
        \includegraphics[width=\linewidth]{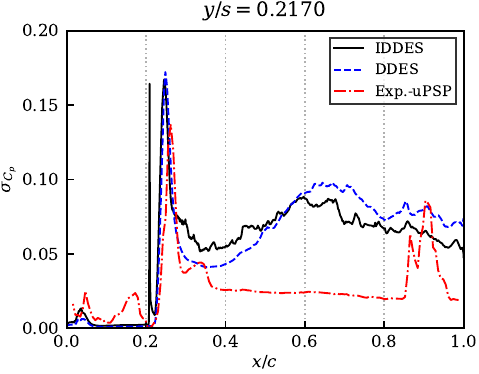}
        \caption{Temporal standard deviation of $C_p$}
        \label{fig:slice_cp_var}
    \end{subfigure}
    \caption{Surface slice at $y/s=0.217$}
    \label{fig:slice_cp}
\end{figure}

To further demonstrate the reliability of the DDES data compared with the more complex IDDES, Figure \ref{fig:slice_cp} shows the mean pressure distribution and the standard deviation along a spanwise cut on the wing lower surface at ($y/s$ = 0.217). The $x$-coordinate is normalized by the local chord length $c$. The location of the shock in the DDES coincides very well with the IDDES shock location (cf Fig. \ref{fig:slice_cp_mean}). Both simulations predict the shock slightly more upstream compared to the wind tunnel experiment. The local pressure minimum located just upstream of the shock also agrees very well with the IDDES data. The region of separated flow downstream of the shock exhibits noticeable differences between the DDES and IDDES results. Only the IDDES simulation reproduces the experimental pressure distribution accurately, although the deviations observed in the DDES case remain moderate. Further downstream, the pressure distributions from both DDES and IDDES converge again. When the shock strength or, equivalently, the pressure rise across the shock is compared with the experimental data, the DDES shows a significantly better agreement than the IDDES.

The standard deviation of the surface pressure along the cut displayed in Fig. \ref{fig:slice_cp_var} also reveals an excellent agreement among the DDES results, the IDDES data\footnote{The attentive reader may notice minor differences in the details of the IDDES data compared with publication \cite{Spinner2025}. These discrepancies arise because the IDDES simulation was continued to generate a longer time series, thereby yielding more fully converged statistics. However, this does not affect any of the conclusions drawn in \cite{Spinner2025} or the present study.}, and the uPSP measurements. In the separated flow region downstream of the shock, the DDES shows the same deviation from the experimental data as the IDDES. Note that the IDDES data exhibits a strong, distinct peak at $x/c\approx 0.2$ that is neither present in the DDES nor the uPSP dataset. This peak stems from the synthetic turbulence generator used only in the IDDES.

\begin{figure}
    \centering
    \includegraphics[width=0.5\linewidth]{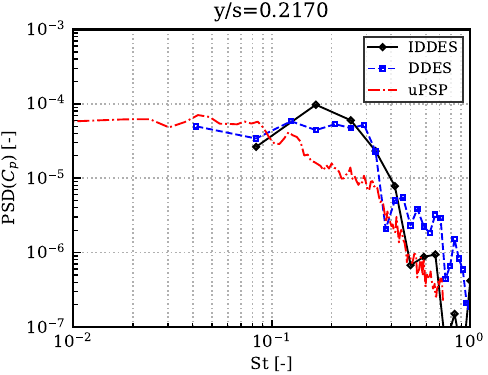}
    \caption{PSD at $\sigma_{C_{p},\mathrm{max}}$ at $y/s=0.217$}
    \label{fig:slice_cp_spectra}
\end{figure}

Further analysis of the unsteady flow phenomena was conducted by extracting the power spectral density (PSD) at the positions of maximum $\sigma_{C_{p}}$ from Fig. \ref{fig:slice_cp_var} ($x/c=0.245$ for IDDES, $x/c=0.248$ for DDES, and $x/c=0.262$ for uPSP). The resulting spectra shown in Fig. \ref{fig:slice_cp_spectra} exhibit good agreement between DDES and IDDES simulations. As can be seen, the frequency resolution could be significantly improved for the DDES simulation compared to the IDDES. Both simulations show a broadband signal in the Strouhal number range of 0.1 to 0.4 with a strong reduction in PSD at $St \approx 0.4$. The experimental data also exhibits these signals, although they are way less pronounced. We performed analogous comparisons at additional spanwise locations, following the methodology of \cite{Spinner2025}, and observed agreements of comparable accuracy to those reported here. For the sake of brevity, the explicit presentation of these results has been omitted.

To summarize, in this and in previous studies \cite{Spinner2024,Spinner2025}, it was found that the PSD spectra of the pressure fluctuations in the area of the shock show increased values in the Strouhal number range of 0.2 to 0.4. To gain a deeper understanding which mechanisms drive these unsteady phenomena, we turn to modal analysis in the following sections.

\subsection{Flow details of shock-induced separation}
\begin{figure}
    \centering
    \begin{subfigure}{0.49\textwidth}
        \includegraphics[width=\linewidth]{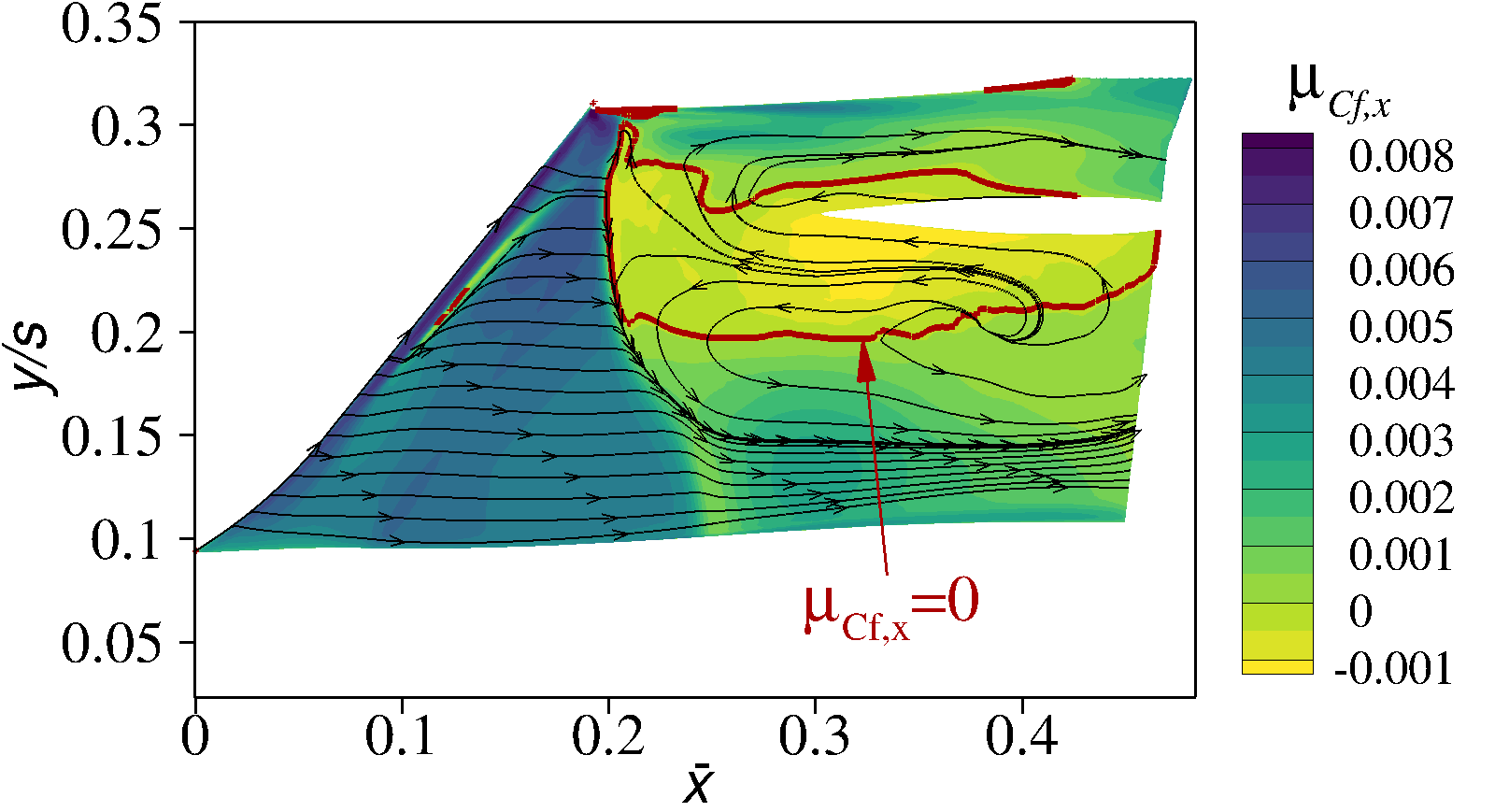}
        \caption{Temporal mean of $C_{f,x}$}
        \label{fig:ddes_meancf}
    \end{subfigure}
    \begin{subfigure}{0.49\textwidth}
        \includegraphics[width=\linewidth]{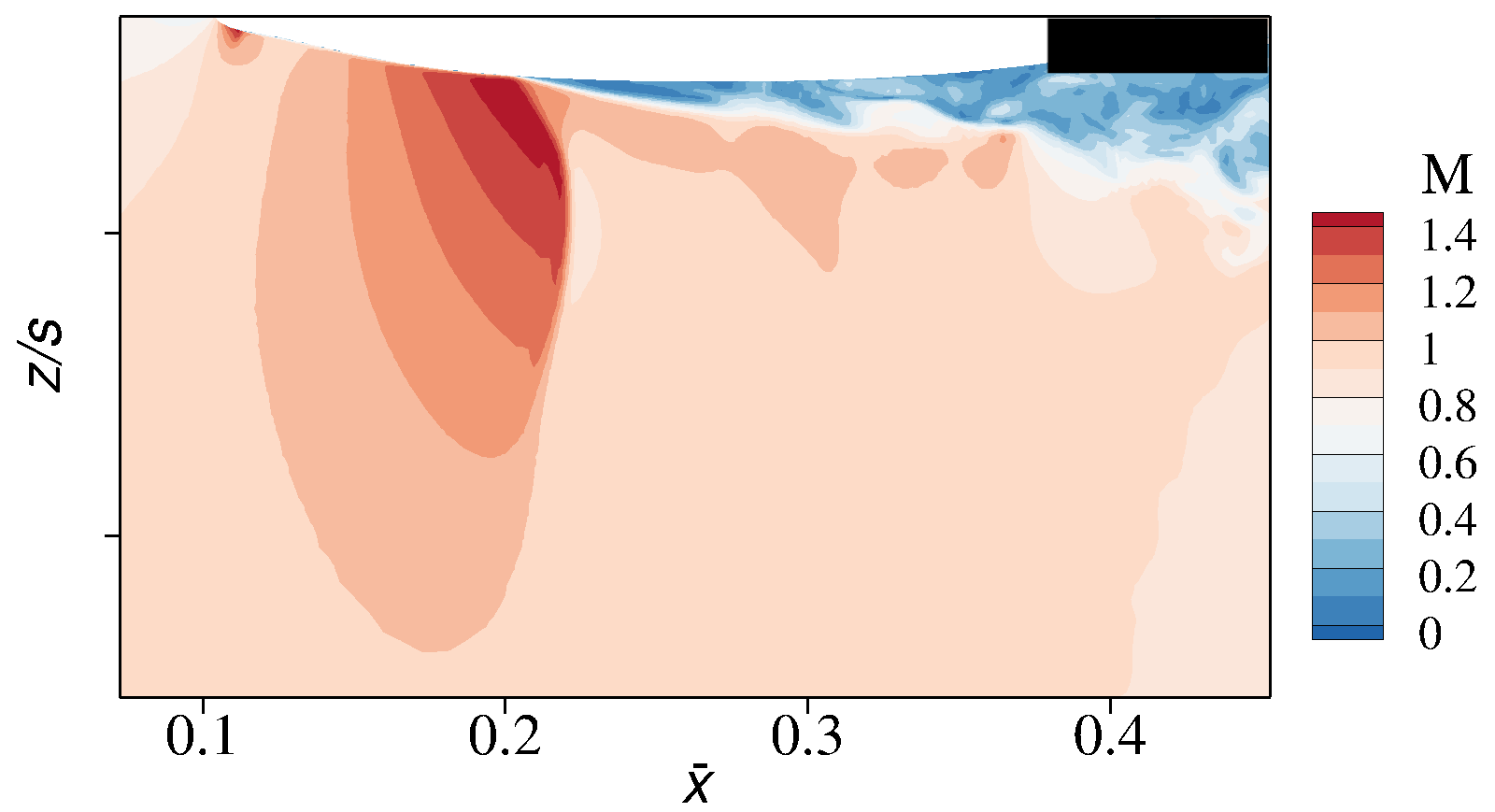}
        \caption{Mach number at $y/s=0.2$, $t_{ref}=50.0$}
        \label{fig:ddes_mach_slice}
    \end{subfigure}
    \caption{Details of flow separation in DDES}
    \label{fig:add_ddes_figs}
\end{figure}

To understand the following modal analysis, it is helpful to consider the flow topology of the shock-induced separation on the lower surface of the wing in addition to the statistical evaluations in the previous chapter. Figure \ref{fig:ddes_meancf} displays the mean skin friction coefficient in $x$-direction ($\mu_{C_{f,x}}$) on the wing surface in the region of interest. The steep drop in wall shear stress across the shock is clearly visible. In addition, the corresponding wall shear stress lines are shown in the figure, and the extension of the flow separation is indicated by a line along $\mu_{C_{f,x}}=0$. The iso-contour reveals the mean spanwise extent of the separation in the range $y/s \in [0.2,0.3]$. It also becomes clear that the separation extends down to the trailing edge and that there are strong local reverse flows.

The instantaneous Mach number field at $t_{ref}=50.0$ ($t_{ref}=t\cdot\mathrm{MAC}/U_{\infty}$ - reference time) on a slice at $y/s=0.2$ is shown in Fig. \ref{fig:ddes_mach_slice} displaying the wall normal extent of the shock-induced flow separation on the wing lower surface (the rear part of the airfoil has been concealed for reasons of confidentiality). The flow separates from the wall immediately downstream of the shock. The wall-normal extent of the flow separation continues to increase downstream. After a certain length downstream of the separation line ($\bar{x}\approx0.28$), the shear layer between the separation region and the surrounding flow breaks down into large-scale turbulent structures.

\subsection{Spectral Proper Orthogonal Decomposition of Surface Snapshots}
SPOD identifies coherent structures that oscillate at specific frequencies, providing a framework for analyzing statistically stationary flows. Each SPOD mode $\boldsymbol{\varphi}$ is a complex-valued vector, and its interpretation requires understanding its four key components: the real part $\Re\left(\boldsymbol{\varphi}\right)$ and the imaginary part $\Im\left(\boldsymbol{\varphi}\right)$ represent the in-phase and quadrature ($90^\circ$ phase-shifted) components of the mode's harmonic oscillation at a selected frequency. Both are equally important and modify the base flow. The magnitude $|\boldsymbol{\varphi}|$ is the most intuitive quantity, directly indicating which flow regions participate in a pattern at what strength.
The local phase angle $\boldsymbol{\phi} = \mathrm{arg}\left(\boldsymbol{\varphi}\right)$ reveals traveling structures, spatial oscillations, or phase lags resulting from spatial variations between the real and imaginary part. Furthermore, traveling structures exhibit smooth phase angle variations, allowing the estimation of propagation direction and speed.
Specifically, the structure propagates in the negative phase gradient direction at a speed inversely proportional to the gradient's magnitude and proportional to the associated frequency.

\begin{figure}
    \centering
    \begin{subfigure}{0.49\textwidth}
        \includegraphics[width=\linewidth]{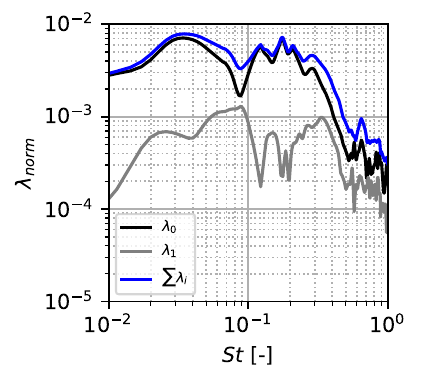}
        \caption{DDES}
        \label{fig:ddes_eig_contrib}        
    \end{subfigure}
    \begin{subfigure}{0.49\textwidth}
        \includegraphics[width=\linewidth]{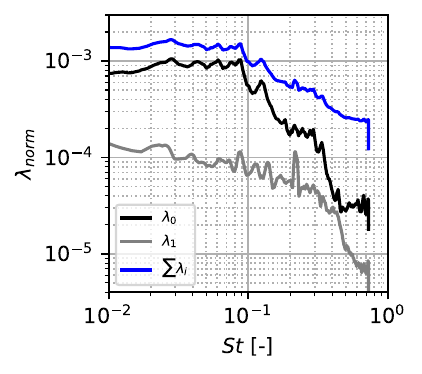}
        \caption{uPSP}
        \label{fig:psp_eig_contrib}        
    \end{subfigure}
    \caption{SPOD eigenvalue contribution.}
    \label{fig:spod_eig_contrib}        
\end{figure}

Figure \ref{fig:spod_eig_contrib} shows the spectra of the first two eigenvalues and the sum over all $N_\mathrm{win}$ eigenvalues at frequency $f_k$, normalized such that $\lambda_{k,i,norm}=\lambda_{k,i}/\sum_{k,i}{\lambda_{k,i}}$.
The relative contribution of an eigenvalue at a specific frequency indicates the importance of the corresponding mode.
If the leading eigenvalue is close to the sum of eigenvalues, then the mode-eigenvalue pair explains most of the variance at that frequency.
The associated flow phenomenon is highly harmonic.
On the other hand, if there are multiple different structures at a given frequency, then the energy is shared among multiple eigenvalues, and the relative contribution is low.
The latter scenario is typical for small-scale turbulence found at the upper end of the spectrum.
Figure \ref{fig:spod_eig_contrib} indicates highly coherent structures in the range $St \in [0.1,0.4]$ of the DDES surface data.
Note that the low-frequency range $St \leq 0.08$ of the DDES spectrum is strongly biased and hence not trustworthy (due to the limited frequency resolution).
Naturally, the leading eigenvalue is less dominant in the more complex uPSP data.
Nonetheless, the trend of the leading eigenvalue closely resembles the (binwise) sum of all eigenvalues, indicating its dominance.
Consequently, we restrict our analysis to the first SPOD mode-eigenvalue pairs at selected frequencies.

\begin{figure}
    \centering
    \includegraphics[width=0.8\textwidth]{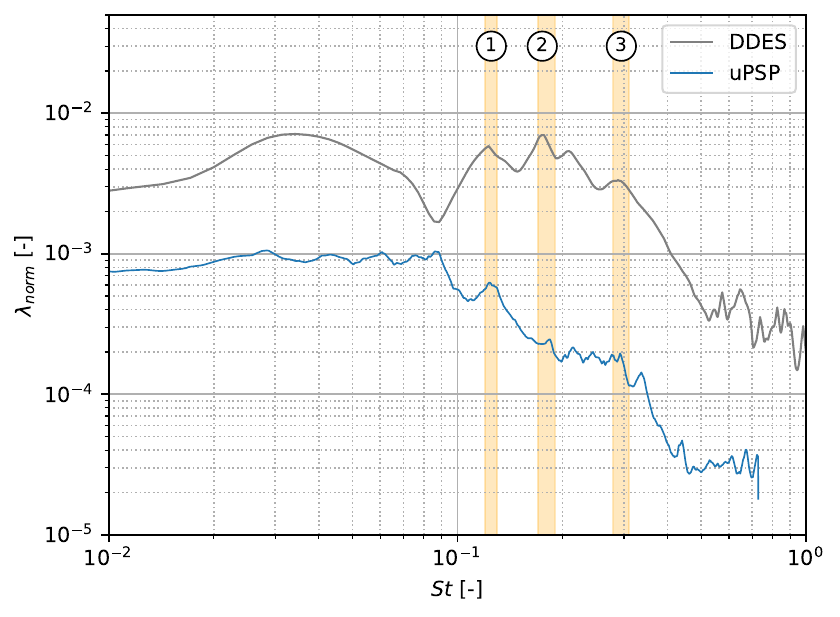}
    \caption{Spectra of the most dominant SPOD eigenvalues.}
    \label{fig:spod_spectra}
\end{figure}

Figure \ref{fig:spod_spectra} shows the resulting SPOD eigenvalue spectra of the most dominant eigenvalue ($\lambda_{k,0}$) for the two surface data sets. While both spectra show significant energy for $St<0.1$, the DDES data is strongly biased and untrustworthy in the range 
 due to limited frequency resolution resulting from only 50 CTU. Additionally, the power spectral density of pressure variation decays rapidly beyond $St>0.4$ for all data sets
 (cf. Fig. \ref{fig:slice_cp_spectra}). This strong decay is also visible in the spectra of the dominant SPOD eigenvalues. Therefore, this study limits the analysis to coherent structures with high modal energy in the Strouhal number range $St \in \left[0.1,0.4\right]$.

Based on Figure \ref{fig:spod_spectra}, we selected three pronounced frequency bands within the buffet range.
They are centered around Strouhal numbers 0.12, 0.18, and 0.29 and numbered \textbf{1} to \textbf{3} in Fig. \ref{fig:spod_spectra} to make them easier to identify. As can be seen, the eigenvalue spectrum of the DDES shows two distinct modes in the vicinity of frequency band \textbf{2} with Strouhal numbers of 0.18 and 0.21, respectively. However, closer inspection showed that these modes are very similar in shape and phase, resulting in redundant information. Therefore, the following analysis will be based on the lower frequency mode of the two. We note that the spectral peaks in the uPSP data are less pronounced; however, for every dominant signal identified in the DDES results, a corresponding (albeit smaller) peak is also present in the uPSP spectrum.
Moreover, the relative contribution of the leading eigenvalue is significantly lower for the uPSP data (the blue curve falls below the gray curve in Fig. \ref{fig:spod_spectra}), indicating a larger rank of the cross-spectral density matrix $\hat{\mathbf{X}}_k\hat{\mathbf{X}}_k^\ast $.
In other words, the spectral changes across different realizations are more pronounced in the experimental data, which is not uncommon and is most likely linked to measurement noise or wind-tunnel effects.
Nonetheless, the relative contribution of the leading eigenvalue is significantly larger than that of the remaining ones for both datasets.

\begin{figure}
    \centering
    \begin{subfigure}{0.49\textwidth}
        \begin{tikzpicture}
            \node (image2) at (0,0) {\includegraphics[width=\linewidth]{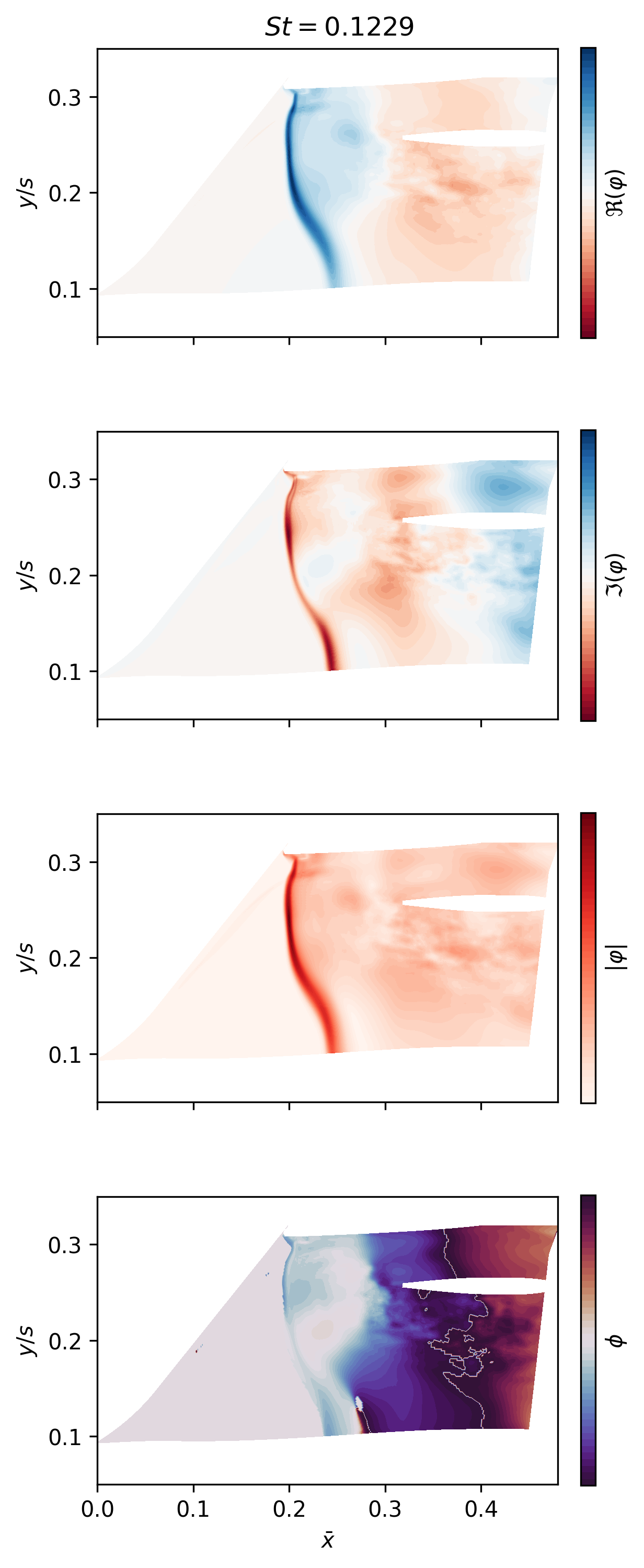}};
            \draw[black, -{Latex}, line width=0.7pt] (-0.2,-5.1) -- (-0.28,-4.6);
            \draw[black, -{Latex}, line width=0.7pt] (-0.18,-5.2) -- (0.0,-5.6);
        \end{tikzpicture}
        \caption{DDES}
        \label{fig:ddes_mode_42}    
    \end{subfigure}
    \begin{subfigure}{0.49\textwidth}
        \includegraphics[width=\linewidth]{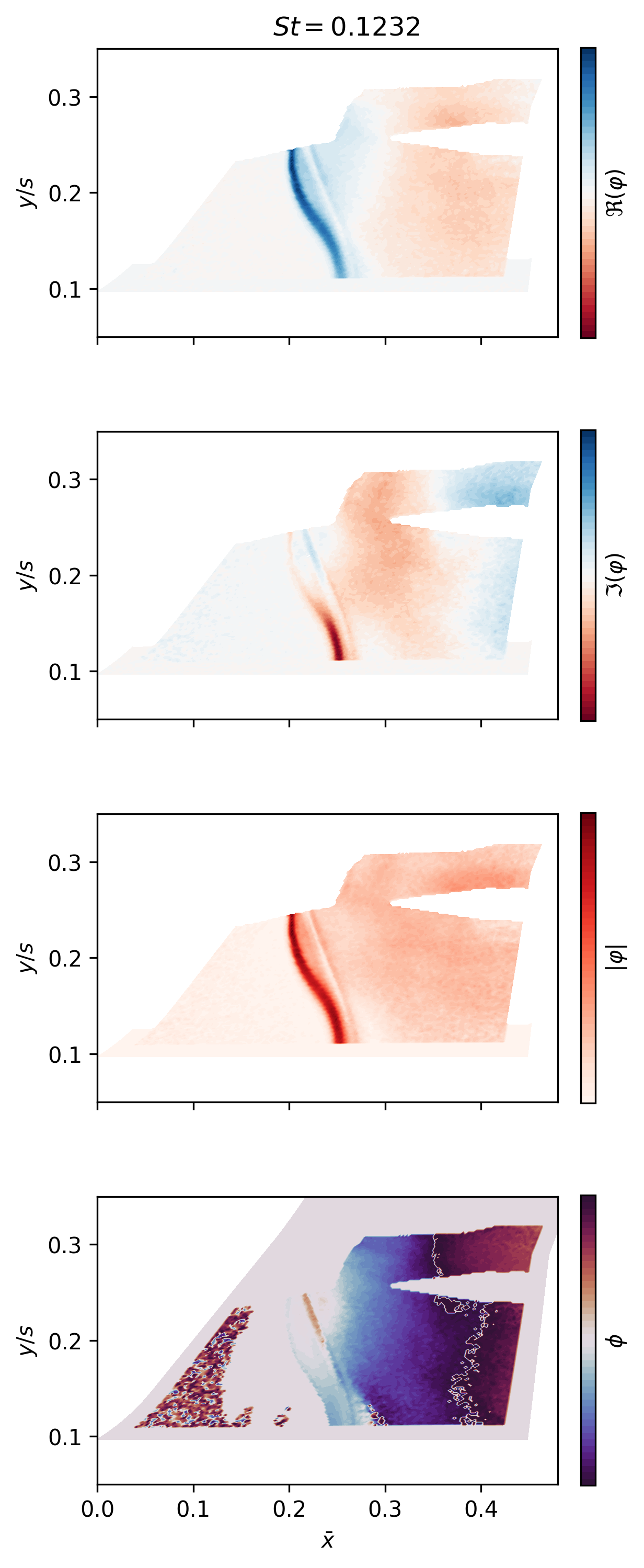}
        \caption{uPSP}
        \label{fig:upsp_mode_296}    
    \end{subfigure}
    \caption{Comparison of dominant SPOD modes in frequency band \textbf{1} at $St \approx 0.12$}
    \label{fig:spod_0.12}
\end{figure}

\begin{figure}
    \centering
    \begin{subfigure}{0.49\textwidth}
        \includegraphics[width=\linewidth]{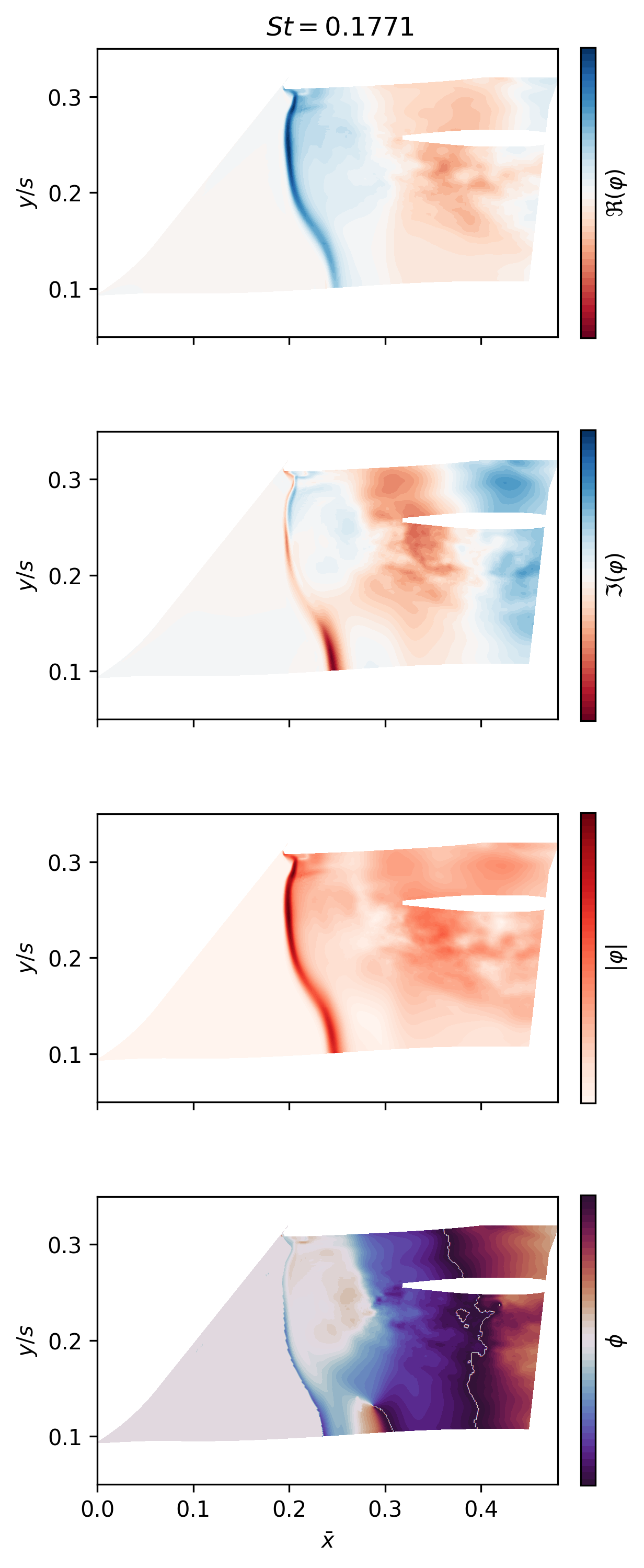}
        \caption{DDES}
        \label{fig:ddes_mode_72}    
    \end{subfigure}
    \begin{subfigure}{0.49\textwidth}
        \includegraphics[width=\linewidth]{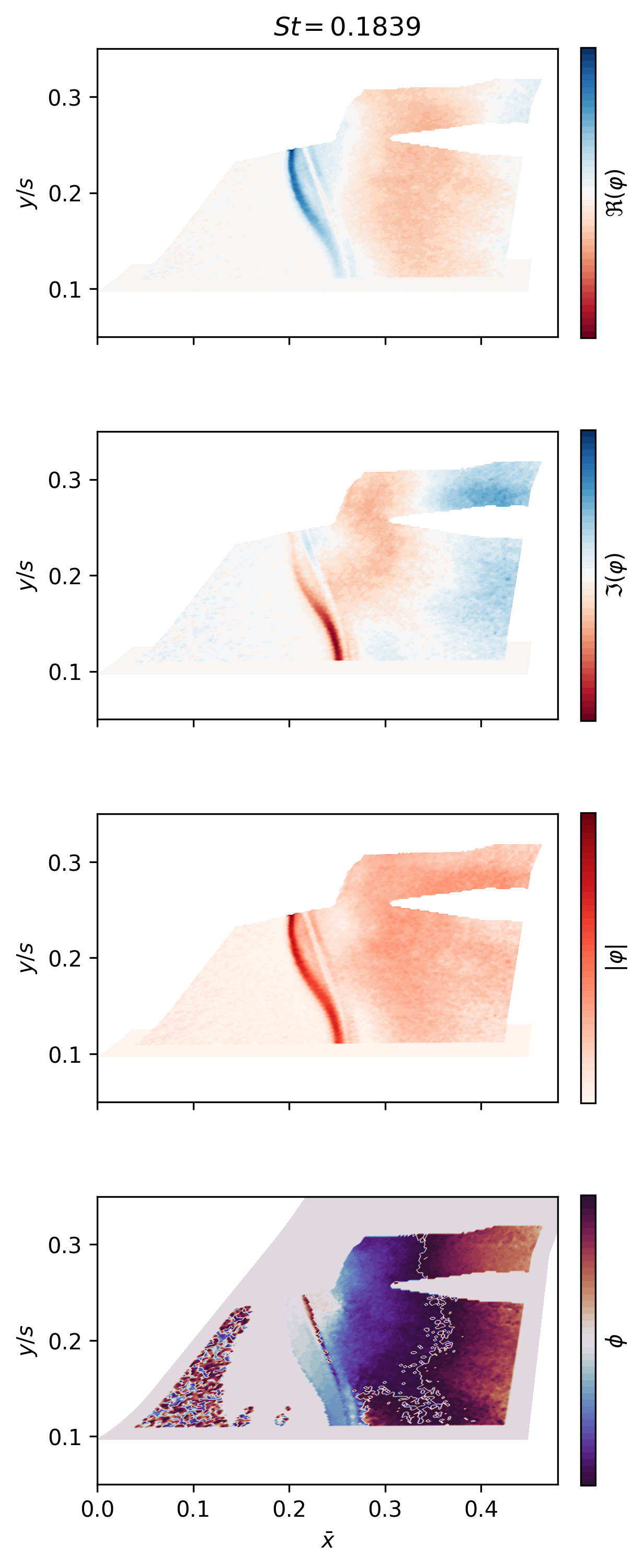}
        \caption{uPSP}
        \label{fig:upsp_mode_515}    
    \end{subfigure}
    \caption{Comparison of dominant SPOD modes in frequency band \textbf{2} at $St \approx 0.18$}
    \label{fig:spod_0.18}
\end{figure}

\begin{figure}
    \centering
    \begin{subfigure}{0.49\textwidth}
        \includegraphics[width=\linewidth]{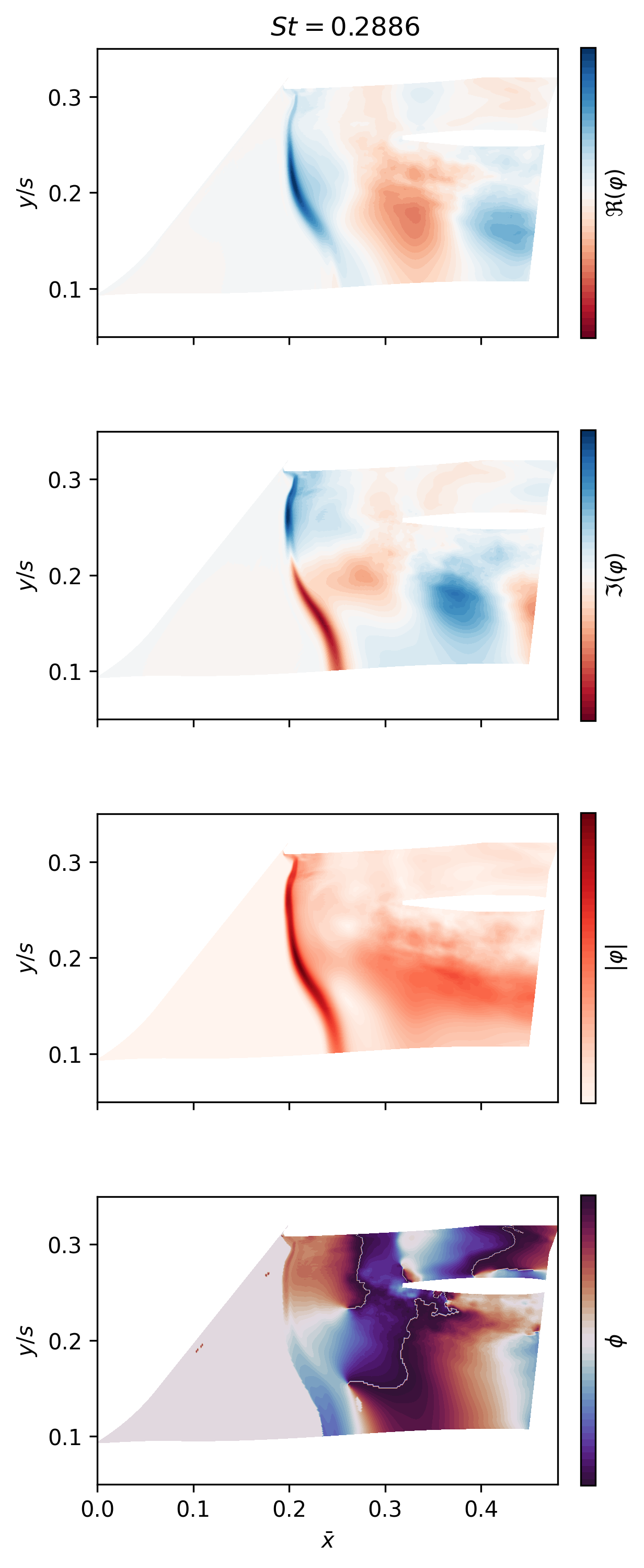}
        \caption{DDES}
        \label{fig:ddes_mode_106}
    \end{subfigure}
    \begin{subfigure}{0.49\textwidth}
        \includegraphics[width=\linewidth]{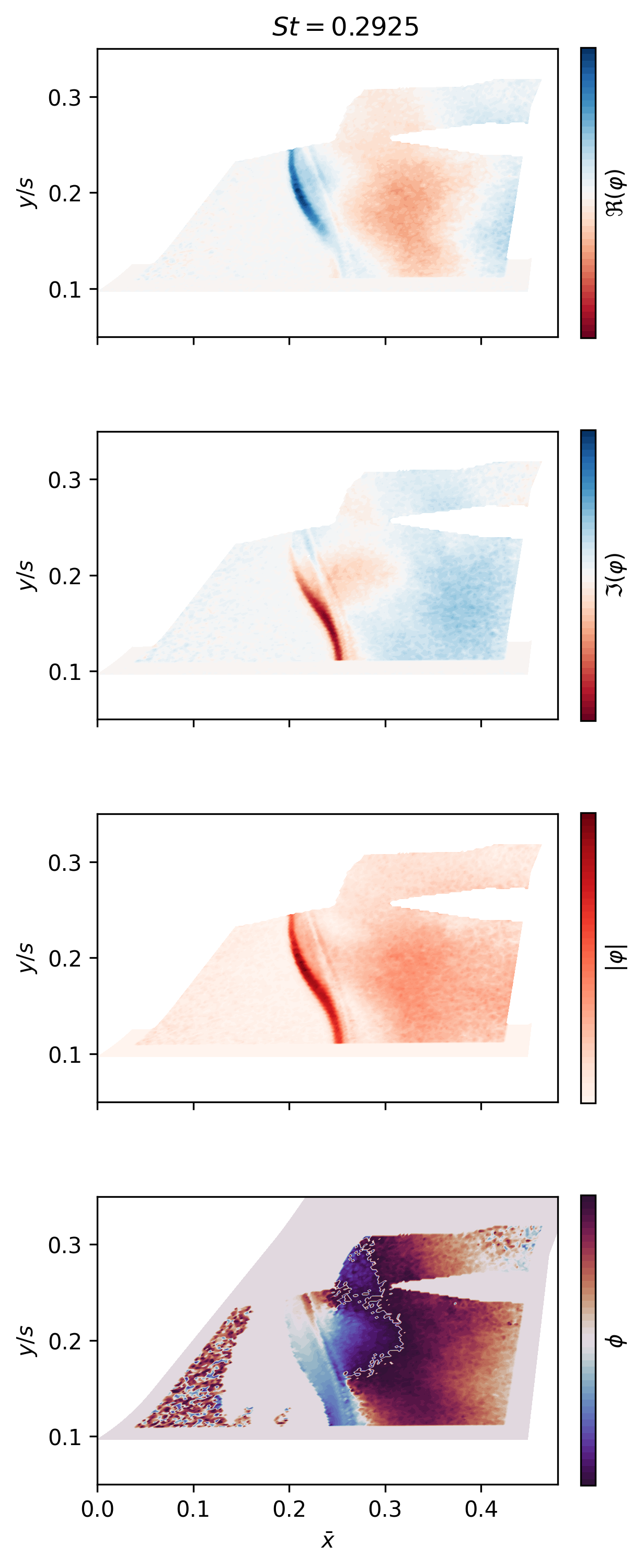}
        \caption{uPSP}
        \label{fig:upsp_mode_703}    
    \end{subfigure}
    \caption{Comparison of dominant SPOD modes in frequency band \textbf{3} at $St \approx 0.29$}
    \label{fig:spod_0.29}
\end{figure}

For each mode $\boldsymbol{\varphi}$, Figures \ref{fig:spod_0.12} to \ref{fig:spod_0.29} show the real part, imaginary part, magnitude, and phase angle. To improve the visual appearance and comparability of the plots, we i) phase-lock the modes based on the reference phase value at $(\bar{x},\,y/s)=(0.22,0.22)$ (approximately the center of shock fluctuations) and ii) zero out regions with low pressure fluctuations in the phase angle plot.
Step i) allows a direct visual comparison using the color scale rather than an indirect comparison based on phase angle change.
Step ii) reduces visual artifacts due to measurement noise in the phase angle plot.

Figure \ref{fig:spod_0.12} shows the SPOD modes of DDES and uPSP in the frequency band \textbf{1} at a Strouhal number of around 0.12. The modes were selected as they are represented by the most dominant eigenvalue in that frequency band.
Given the modes' magnitudes, the base flow is mainly modified in the shock region and to a lesser extent downstream of it.
The real and imaginary parts show how these modifications are correlated.
The previously described alignment of the modes enables a direct comparison of the real and imaginary parts of the modes across the two datasets.
We note the close resemblance between $\Re(\boldsymbol{\varphi}_1^\mathrm{DDES})$ and $\Re(\boldsymbol{\varphi}_1^\mathrm{uPSP})$ and the similarity between $\Im(\boldsymbol{\varphi}_1^\mathrm{DDES})$ and $ \Im(\boldsymbol{\varphi}_1^\mathrm{uPSP})$.
In both datasets, pressure changes downstream of the shock alternate between positively and negatively correlated portions of the flow.
This modulation indicates a flow separation near the shock and a subsequent convective transport of the pressure disturbance.
The propagation of the pressure disturbance with the mean flow is confirmed by the phase angle slope (the phase angle decreases in the chordwise direction).
For the DDES mode, the propagation of disturbances in the area of the shock appears to originate at approximately $y/s=0.2$. From there, the phase angle distribution indicates propagation along the shock, both inboard and outboard (the phase angle reduces in both directions along the shock as indicated by the two arrows). Further studies suggest that the origin of these oscillations coincides with the spanwise transition between the detached and attached boundary layer behind the shock (cf. Fig. \ref{fig:ddes_meancf}). As visible in the mean skin friction plot in Fig. \ref{fig:ddes_meancf}, the inboard reattachment of the flow downstream of the shock begins in the area of $0.2>y/s>0.15$.
The spanwise motion of the DDES mode is evident in the shock region and behind it up to $\bar{x}\approx0.3$. In this region, a spanwise motion of the inboard boundary of the separation bubble can be observed in animations of the original data (see supplementary material).
Unfortunately, we can not confirm this phenomenon in the uPSP mode analysis, as the region where the outboard propagation would occur is not visible in the uPSP image database. However, since the visible part of the uPSP mode agrees very well with the DDES, we suspect the phenomenon is also present in the wind tunnel experiment.

When examining the uPSP mode, a secondary structure appears immediately downstream of the shock as a bright band in the $\Re(\boldsymbol{\varphi}_1)$, $\Im(\boldsymbol{\varphi}_1)$, and $|\boldsymbol{\varphi}_1|$ fields. This feature is not an aerodynamic phenomenon but a measurement artifact. In this region, the line of sight of the PSP camera intersected the transonic shock, and the strong density gradient caused a local change in the refractive index of the fluid. The resulting optical distortion manifests in the modal analysis of the uPSP images as the observed bright band. However, further analysis revealed that this artifact has no significant impact on the modal analysis.

The dominant modes of frequency band \textbf{2} centered around $St=0.18$ are shown in Fig. \ref{fig:spod_0.18}. The magnitude plot of both modes $\boldsymbol{\varphi}_2^{DDES}$ and $\boldsymbol{\varphi}_2^{uPSP}$ indicates that the base flow is mainly modified by these modes in the shock region and downstream of it, similar to the observations made for frequency band \textbf{1}.
Considering how these modifications are correlated, both real and imaginary parts agree very well between DDES and uPSP. We observe alternating positive and negative pressure correlations downstream of the shock as well as a change in pressure correlation along the spanwise extent of the shock.
The modes show a decreasing phase angle $\boldsymbol{\phi}$ along the shock with decreasing $y/s$. This indicates coherent wave-like structures moving from the area around the pylon, along the shock on the wing lower surface, towards the fuselage. The unsteady behaviour in the shock region described by this mode may be divided into two areas. For the shock region $y/s>0.2$, the phase angle $\boldsymbol{\phi}_2^{DDES}$ is nearly constant, indicating a homogeneous pulsation of the shock. For $\boldsymbol{\phi}_2^{uPSP}$ a similar behavior is indicated in the visible region. Considering the superposition of this mode and the base flow, the pulsation effectively describes a shock oscillation in the chordwise direction. Further inboard, $y/s<0.2$, the phase shift indicates an additional propagation of the shock oscillation towards the fuselage.
For the region of separated flow downstream of the shock, we observe a negative phase angle gradient with increasing $\bar{x}$ for the DDES and the uPSP mode, indicating downstream propagation of separated flow structures.
As described in the preceding analysis, an optical artifact is evident in the uPSP data slightly downstream of the shock.

The SPOD modes of frequency band \textbf{3} at $St \approx 0.29$ are shown in Fig. \ref{fig:spod_0.29}. As for the previously discussed modes, we observe high mode magnitudes in the area of the shock spanning over the entire visible range of $y/s$. In the DDES data, the regions downstream of the shock and inboard of the flap track fairing show significantly elevated magnitudes, especially for $\bar{x}>0.28$, which was not observed in the previously discussed modes. The uPSP mode thereby agrees very well with the DDES mode, exhibiting a smooth region of increased magnitudes between the fuselage and the flap track fairing for $\bar{x}>0.28$. 
In contrast to the low‑frequency modes examined previously, the structures observed in $\Re(\boldsymbol{\varphi}_3)$ and $\Im(\boldsymbol{\varphi}_3)$ exhibit significantly reduced wavelengths. Downstream of the shock, alternating pressure correlations appear not only in the streamwise direction, but also in the spanwise direction. Moreover, the streamwise wavelength of these alternating pressure correlations is considerably shorter than observed in frequency bands \textbf{1} and \textbf{2}.
The phase angle distribution along the shock indicates structures propagating from the pylon-wing-intersection towards the fuselage. Both modes exhibit a negative phase angle gradient in the separation region downstream of the shock, indicating propagation of coherent structures towards the trailing edge. On closer inspection, it becomes apparent that these downstream propagating disturbances appear to originate downstream of the shock in the region $y/s\approx0.22$ and are then transported slightly diagonally (decreasing $y/s$) towards the trailing edge.

Overall, the analysis reveals wave-like disturbances propagating along the shock front toward the fuselage, similar to buffet cells in classical shock buffet. While the phenomena overlap across modes, distinct characteristics emerge: the mode in frequency band \textbf{1} incorporates a spanwise "breathing" motion of the downstream flow separation; the mode in frequency band \textbf{2} features 2D-like shock oscillations where the shock is perpendicular to the flow, transitioning to spanwise-convecting buffet cells where the shock becomes oblique; the mode in frequency band \textbf{3} includes higher-order buffet cells with vortex shedding effects from shear layer breakup. Downstream of the shock, coherent structures propagate toward the trailing edge and remain phase-linked to shock movements across all modes. Crucially, every mode exhibits shock oscillations and associated downstream structure propagation, demonstrating that shock oscillations are closely linked to the downstream boundary layer unsteadiness.
Moreover, the buffet phenomenon cannot be attributed to a single mode or phenomenon.

\subsection{Spectral Proper Orthogonal Decomposition of Volume Snapshots}
In addition to the surface mode analysis presented in the previous chapter, the volumetric data generated by the DDES enable a spatial modal analysis of the buffet phenomenon.
Owing to computational constraints, the snapshots were stored at a temporal resolution that is five times coarser than that used for the surface data, thereby reducing the frequency resolution of the volumetric analysis relative to the surface pressure analysis. Despite this limitation, the SPOD spectrum of the volumetric data still revealed dominant signals at frequencies that are virtually identical to those identified in the surface analysis.

\begin{figure}
    \centering
    \begin{subfigure}{0.49\textwidth}
        \centering
        \begin{tikzpicture}
            \node (image1) at (0,0) {\includegraphics[width=\linewidth]{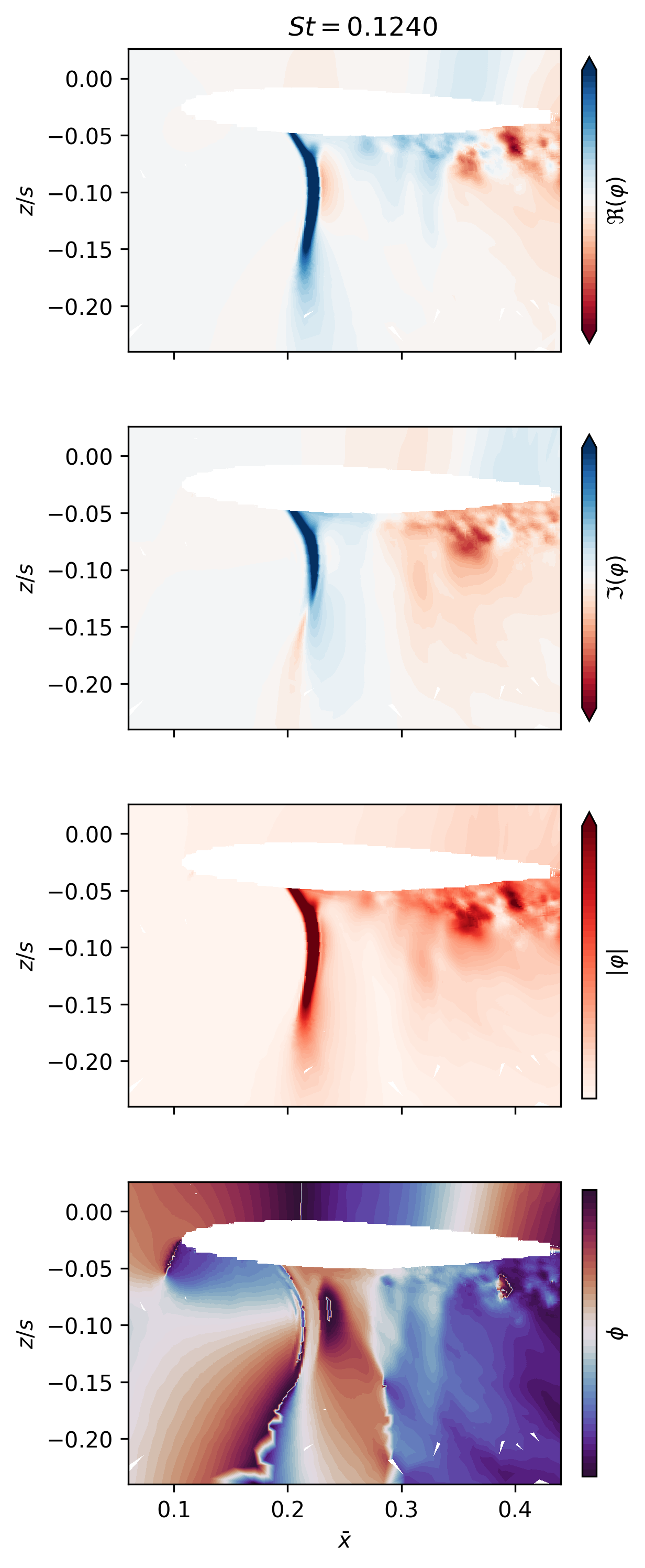}};
            \draw[black, -{Latex}, line width=1pt] (1.7,-4.0) -- (0.7,-3.9);
            \draw[black, -{Latex}, line width=1pt] (0.2,-4.8) -- (1.2,-4.75);
            \draw[black, -{Latex}, line width=1pt] (0.5,-6.0) -- (-0.3,-6.0);
            \draw[black, -{Latex}, line width=1pt] (-0.75,-5.8) -- (-1.3,-5.2);
        \end{tikzpicture}
        \caption{frequency band \textbf{1}}
        \label{fig:spod_vol_0.12}
    \end{subfigure}
    \begin{subfigure}{0.49\textwidth}
        \centering
        \begin{tikzpicture}
            \node (image2) at (0,0) {\includegraphics[width=\linewidth]{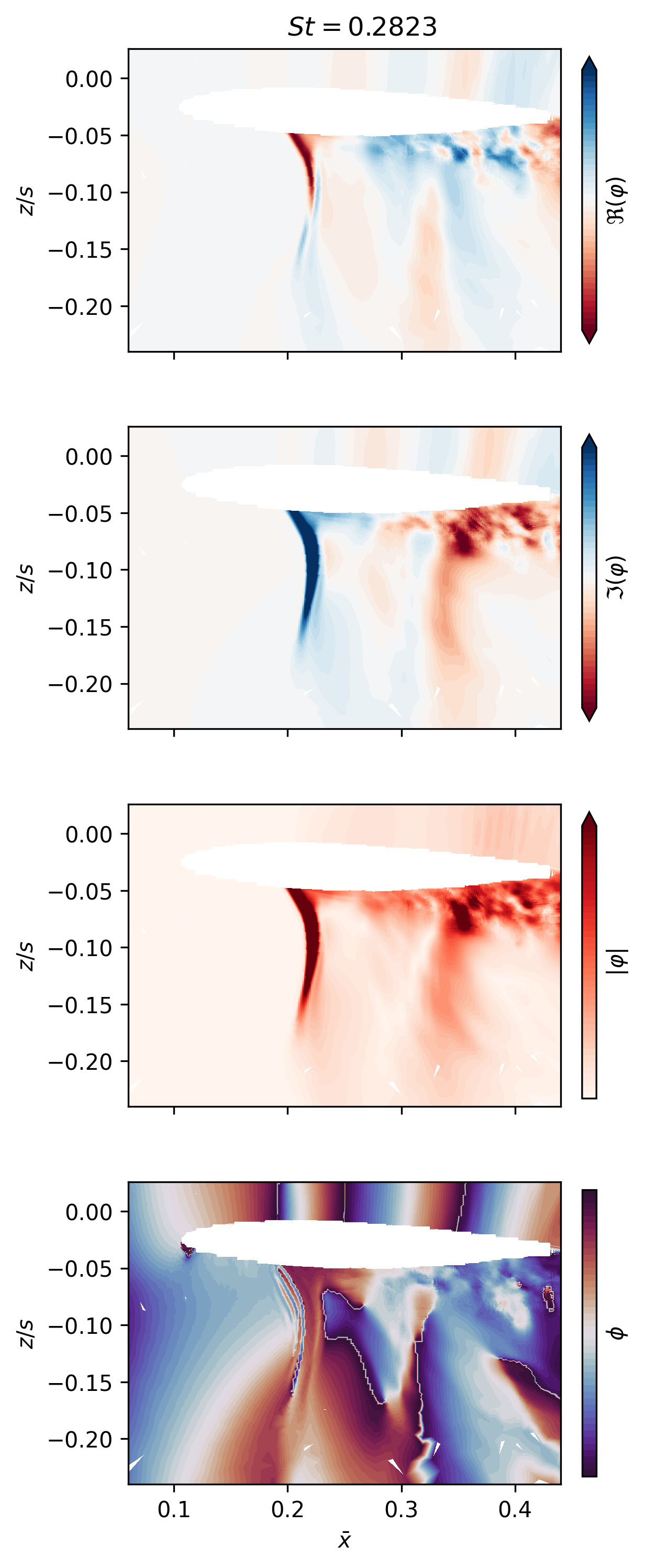}};
            \draw[black, -{Latex}, line width=1pt] (1.6,-4.0) -- (0.6,-3.9);
            \draw[black, -{Latex}, line width=1pt] (0.2,-4.65) -- (1.2,-4.6);
            \draw[black, -{Latex}, line width=1pt] (1.6,-5.5) -- (1.0,-6.0);
            \draw[black, -{Latex}, line width=1pt] (0.6,-5.0) -- (0.0,-5.7);
            \draw[black, -{Latex}, line width=1pt] (-0.7,-6.05) -- (-1.3,-5.4);
        \end{tikzpicture}
        \caption{frequency band \textbf{3}}
        \label{fig:spod_vol_0.29}
    \end{subfigure}
    \caption{Slice plot of SPOD modes at $y/s=0.2$ (airfoil geometry modified, non-representative of XRF1)}
    \label{fig:spod_vol_slices}
\end{figure}

Figure \ref{fig:spod_vol_0.12} represents a slice of the leading SPOD mode in frequency band \textbf{1} extracted at $y/s=0.2$ and matches the surface mode shown in Fig. \ref{fig:spod_0.12}. However, note that the surface and volume modes are not phase-aligned. In line with the previous section, the figure displays real part, imaginary part, magnitude, and phase angle of the mode. To aid the analysis, important propagation directions inferred from the phase angle gradient are marked with black arrows in the phase angle plot.
From the analysis of the real and imaginary components, along with the mode's magnitude, clear correlations emerge between pressure fluctuations along the shock and in the rear portion of the airfoil. The pressure variation across the shock (blue in $\Re(\boldsymbol{\varphi}_1)$) is thereby inversely correlated with the pressure variation in the field near the trailing edge (red in $\Re(\boldsymbol{\varphi}_1)$). The phase angle plot in Fig. \ref{fig:spod_vol_0.12} indicates that structures near the wall, located behind the transonic shock, propagate towards the trailing edge. This can be interpreted as vortical structures emerging from the shock-induced separation, which are convected downstream towards the trailing edge. Both the inverse correlation between shock and trailing edge as well as the downstream propagation were also observed in the corresponding surface mode (cf. Fig. \ref{fig:ddes_mode_42}), supporting this interpretation.

The shock itself exhibits a negative phase angle gradient normal to the wall. This allows us to interpret that the shock oscillation first occurs close to the wall, and the rest of the shock follows further away from the wall with a certain phase lag.

Weak structures are also evident on the rear upper surface of the wing. Analysis of the real and imaginary components, in conjunction with the phase angle gradients, indicates the presence of waves propagating upstream against the mean flow in this region. The identification of these features as acoustic waves is further supported by propagation velocities estimated from the SPOD phase gradients, which are consistent with the local speed of sound.

Similar structures are also observed in the region of the shock on the lower surface of the wing, away from the wall. As illustrated in the slice shown in Fig. \ref{fig:spod_vol_0.12}  at $\bar{x}\in[0.2,0.28]$ and $z/s\le‑0.15$, upstream propagating waves can be identified within the volume. 
These waves then extend in negative x-direction past the shock, with their propagation direction shifting obliquely into the supersonic region, ultimately propagating in the direction of the airfoil’s leading edge. 
However, it is noted that the mode's magnitude in this region is almost negligible, and therefore, the flow is not expected to be significantly influenced by this phenomenon.

As another example of investigating mode propagation in the volume, a mode in frequency band \textbf{3} was selected, matching the surface mode shown in Fig. \ref{fig:spod_0.29}. Figure \ref{fig:spod_vol_0.29} presents a slice view of this mode at $y/s=0.2$. As before, surface and volume modes are not phase-aligned. A separate illustration of the volumetric mode in frequency band \textbf{2} is not included, as it exhibits similar topological features as $\boldsymbol{\varphi}_3$.
However a reconstruction of that mode is shown at the end of this chapter.

Upon examining the real part, imaginary part, and magnitude of the mode, structures similar to those observed in frequency band \textbf{1} become apparent. The mode exhibits a strong influence in the region of the shock and near the wall in the rear part of the airfoil. The spatial wavelengths of the mode's features are significantly shorter compared to the mode in frequency band \textbf{1}.

Downstream of the shock, close to the wall, the phase angle gradient again indicates downstream propagating disturbances of separated flow structures in line with the surface mode analysis in Fig. \ref{fig:ddes_mode_106}. Especially for $\Im(\boldsymbol{\varphi}_3)$ we observe a strong inverse correlation of pressure in the shock region and below the rear part of the wing for $\bar{x}>0.3$.
Furthermore, it can be observed that the magnitude in this area remains relatively high even far away from the wall up to $z/s\approx-0.2$. In this region, the phase angle gradient indicates a propagation direction in negative x-direction and away from the wall (two parallel black arrows in Fig. \ref{fig:spod_vol_0.29}). Wave-like patterns in this region can also be observed in the mode's magnitude. Further studies suggest that these disturbances are related to the breakdown of the shear layer between the separation bubble and the surrounding flow. This can also be seen in the volume slice through the Mach number field in Fig. \ref{fig:add_ddes_figs}. Directly downstream of the shock, the shear layer at the edge of the shock-induced separation is stable before it breaks down into large-scale turbulent structures further downstream. These vortices then generate pressure waves, which become noticeable in the mode starting from $\bar{x}\approx0.25$. These volumetric shear layer structures are also assumed to affect the surface pressure fluctuations observed in the corresponding surface modes (cf. Figs. \ref{fig:spod_0.12}-\ref{fig:spod_0.29}), confirming the coupling between volume dynamics and surface buffet.

As in the analysis of frequency band \textbf{1}, upstream acoustic waves can also be observed in the rear area of the upper side of the wing. In this case, they are even more pronounced. The upstream disturbances on the lower side far away from the wall ($z/s<-0.1$) are also deflected at the shock in the same way observed before, so that for $\bar{x}<0.2$ they run diagonally into the supersonic region towards the leading edge. As before, the mode's magnitude in this area is negligible.
Although not shown in the figure, forward-propagating acoustic waves are detected at many other spanwise locations in both frequency band \textbf{2} and frequency band \textbf{3}.

\begin{figure}[htbp]
    \centering
    
    \begin{subfigure}[b]{0.48\textwidth}
        \centering
        \scalebox{1}[-1]{\includegraphics[width=\linewidth]{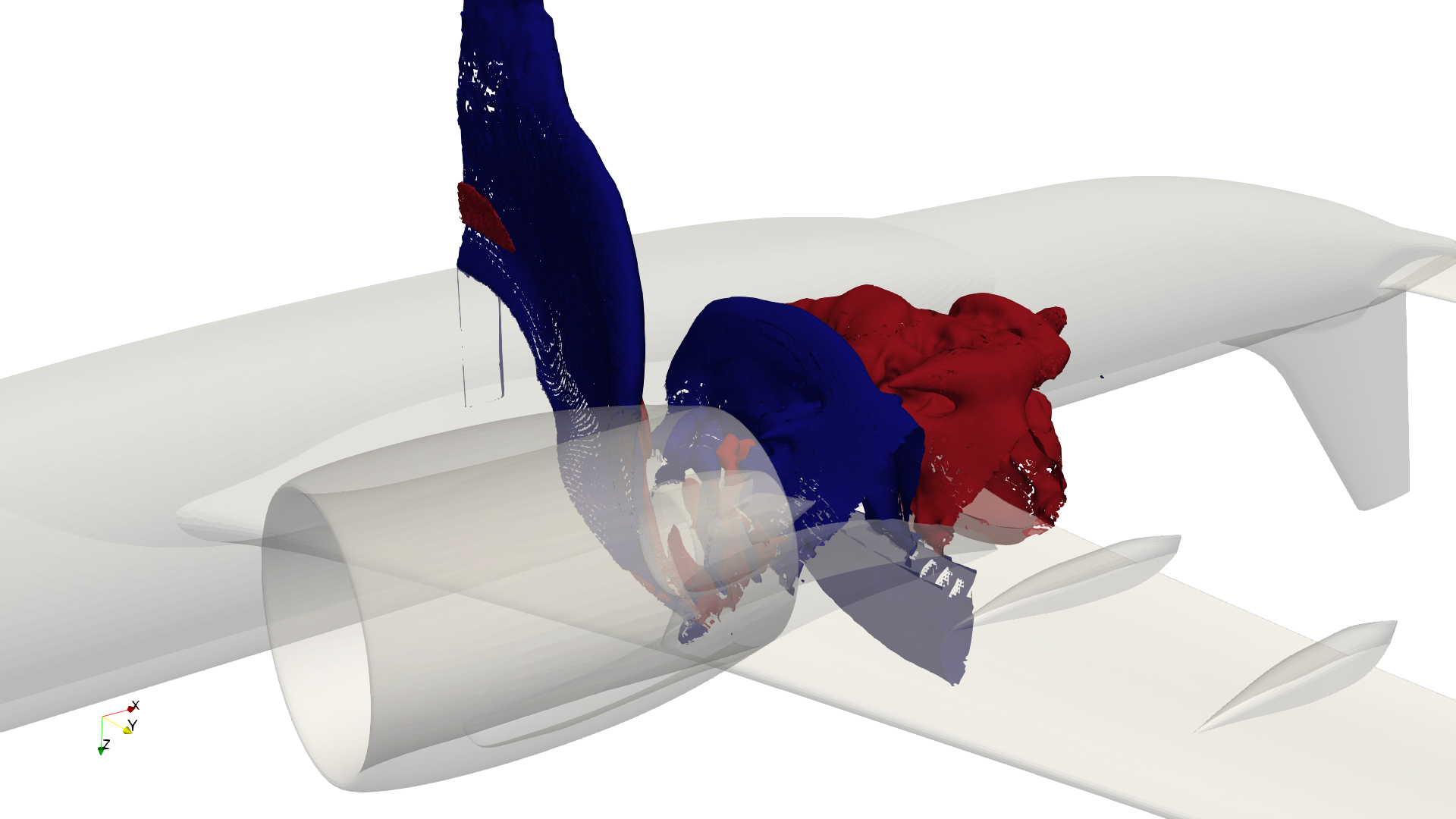}}
        \caption{$\tau=0$}
       \label{fig:sub1}
    \end{subfigure}
    \hfill
    \begin{subfigure}[b]{0.48\textwidth}
       \centering
        \scalebox{1}[-1]{\includegraphics[width=\linewidth]{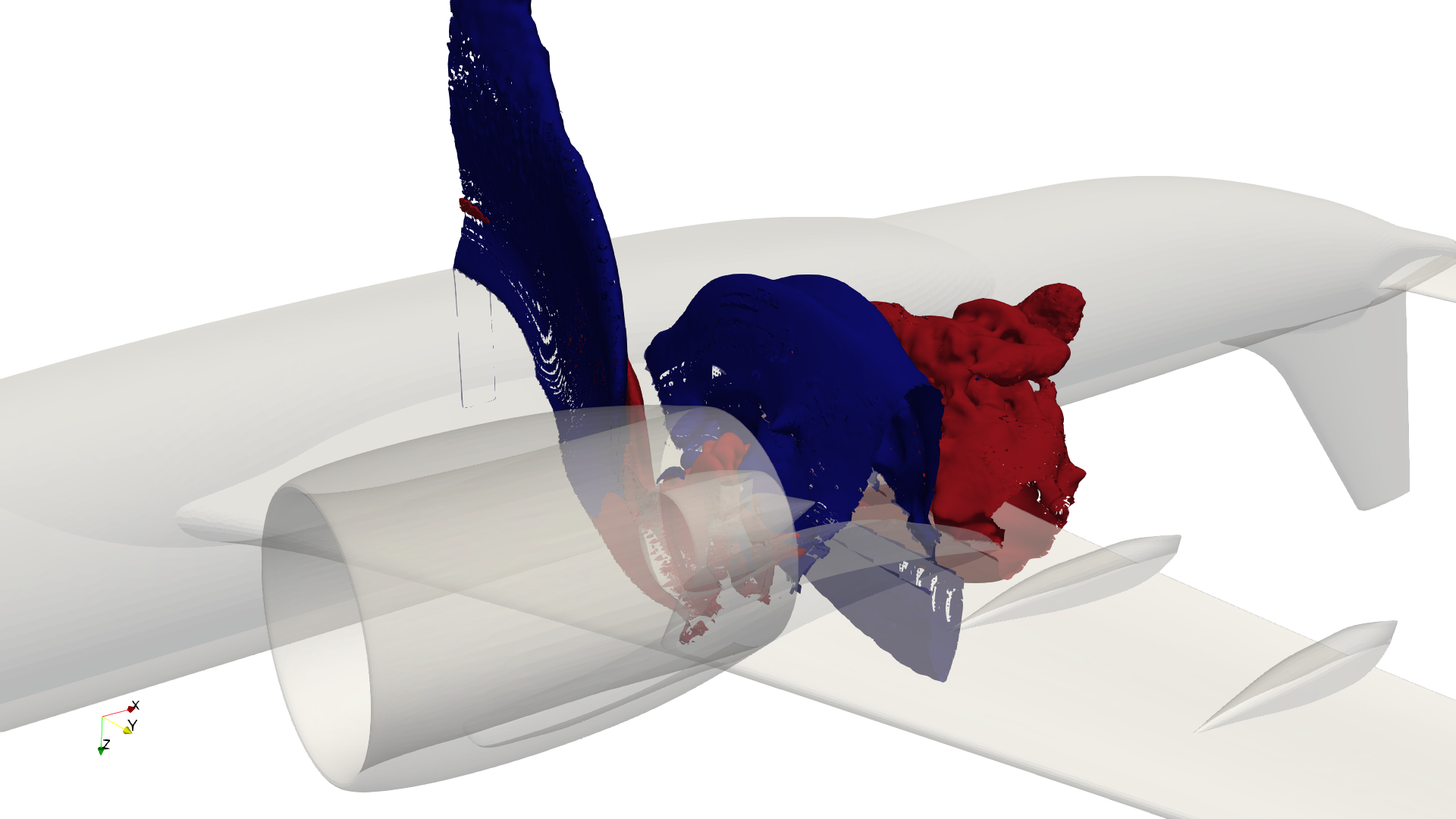}}
        \caption{$\tau=0.1$}
        \label{fig:sub2}
    \end{subfigure}

    \vspace{2ex}

    \begin{subfigure}[b]{0.48\textwidth}
        \centering
        \scalebox{1}[-1]{\includegraphics[width=\linewidth]{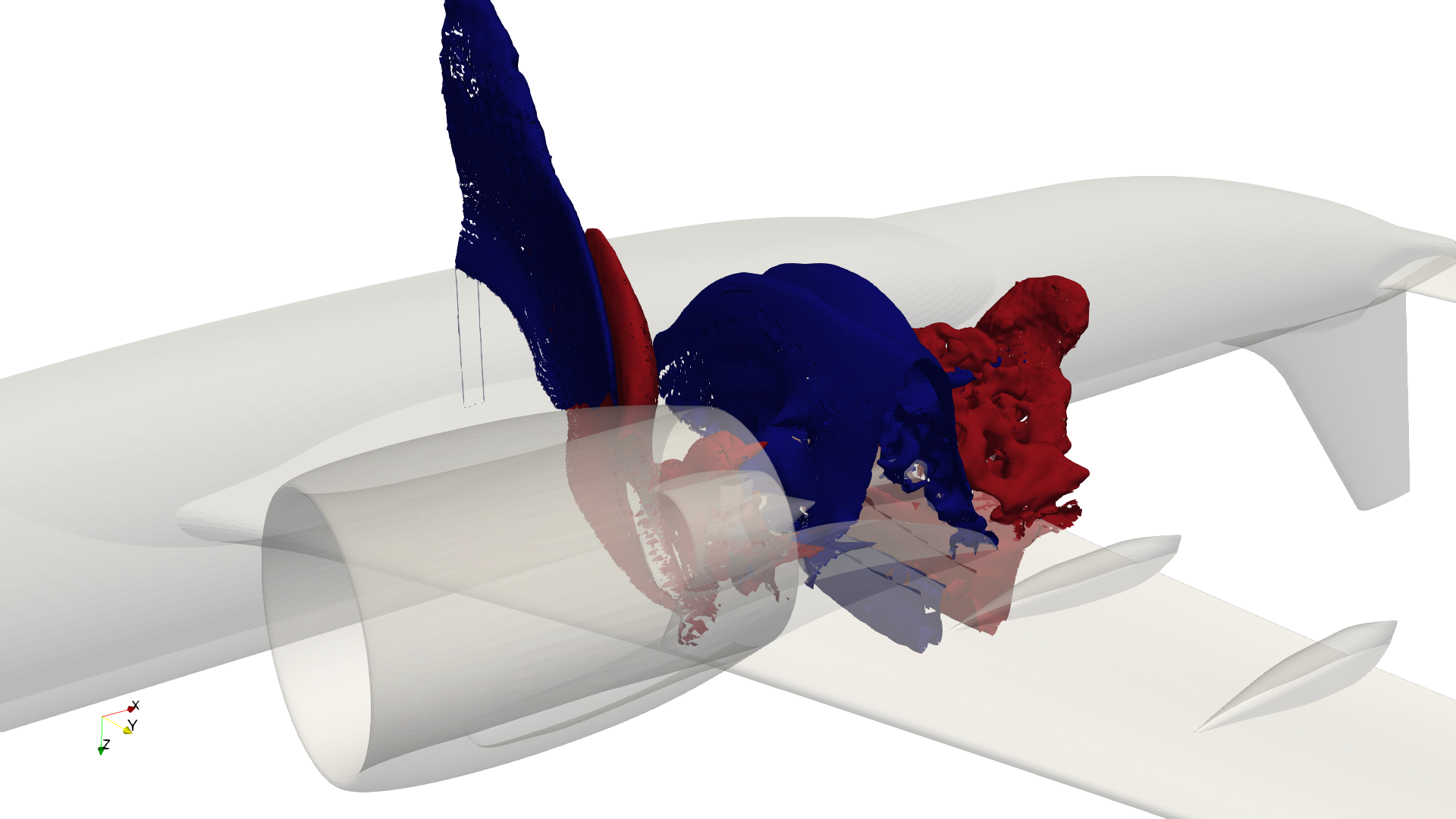}}
        \caption{$\tau=0.2$}
        \label{fig:sub3}
    \end{subfigure}
    \hfill
    \begin{subfigure}[b]{0.48\textwidth}
        \centering
        \scalebox{1}[-1]{\includegraphics[width=\linewidth]{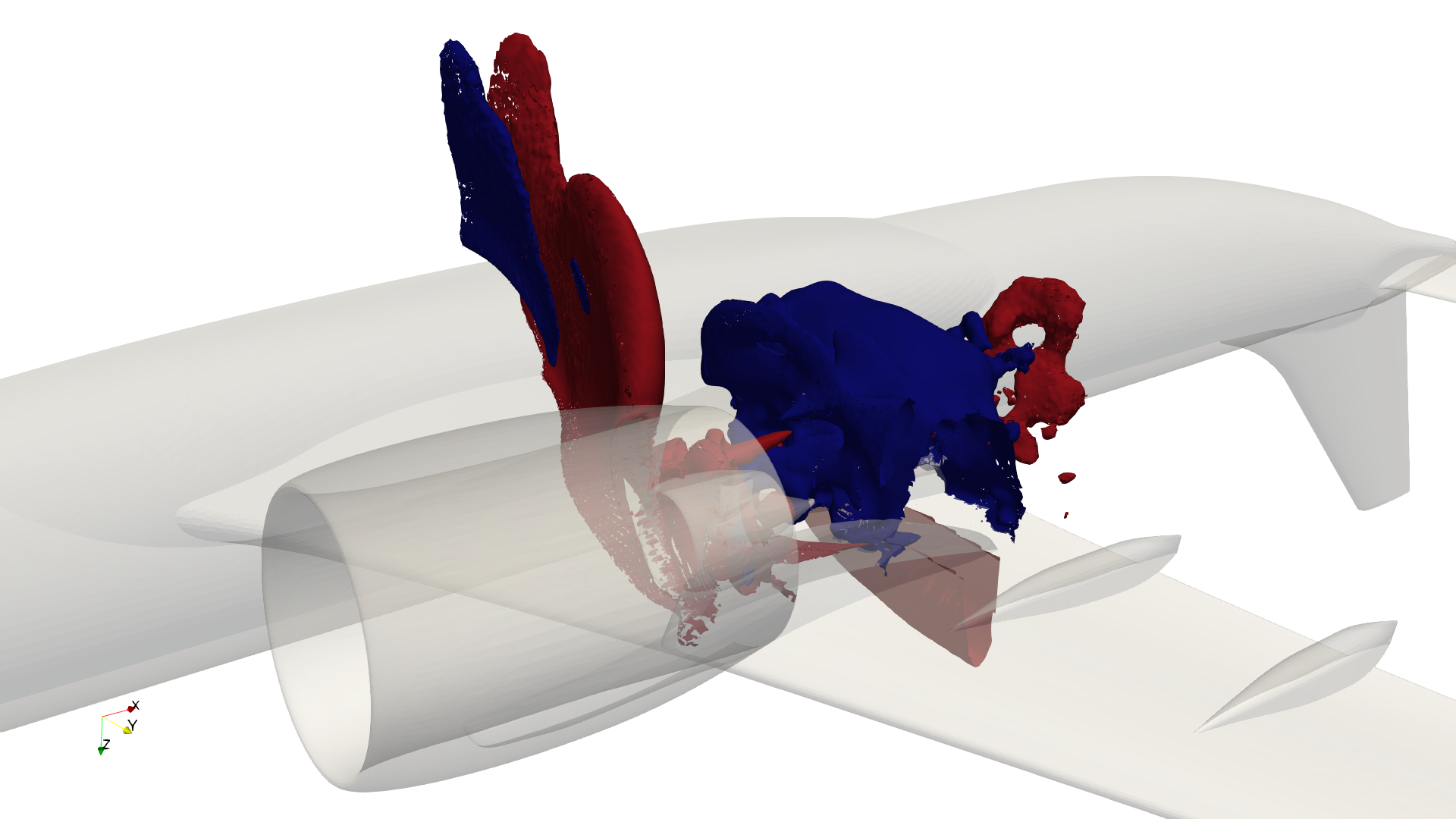}}
        \caption{$\tau=0.3$}
        \label{fig:sub4}
    \end{subfigure}

    \vspace{2ex} 

    \begin{subfigure}[b]{0.48\textwidth}
        \centering
        \scalebox{1}[-1]{\includegraphics[width=\linewidth]{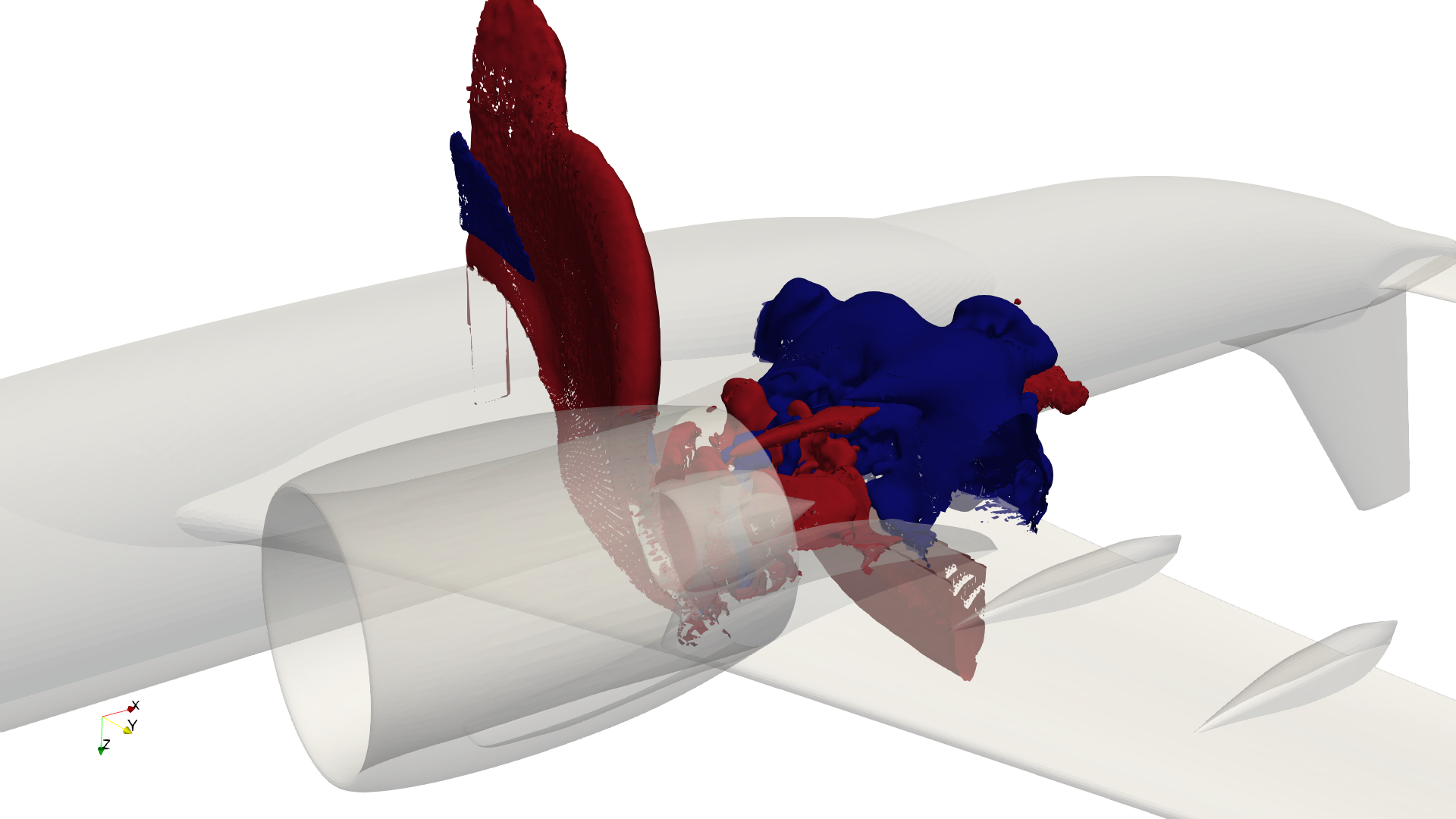}}
        \caption{$\tau=0.4$}
        \label{fig:sub5}
    \end{subfigure}
    \hfill
    \begin{subfigure}[b]{0.48\textwidth}
        \centering
        \scalebox{1}[-1]{\includegraphics[width=\linewidth]{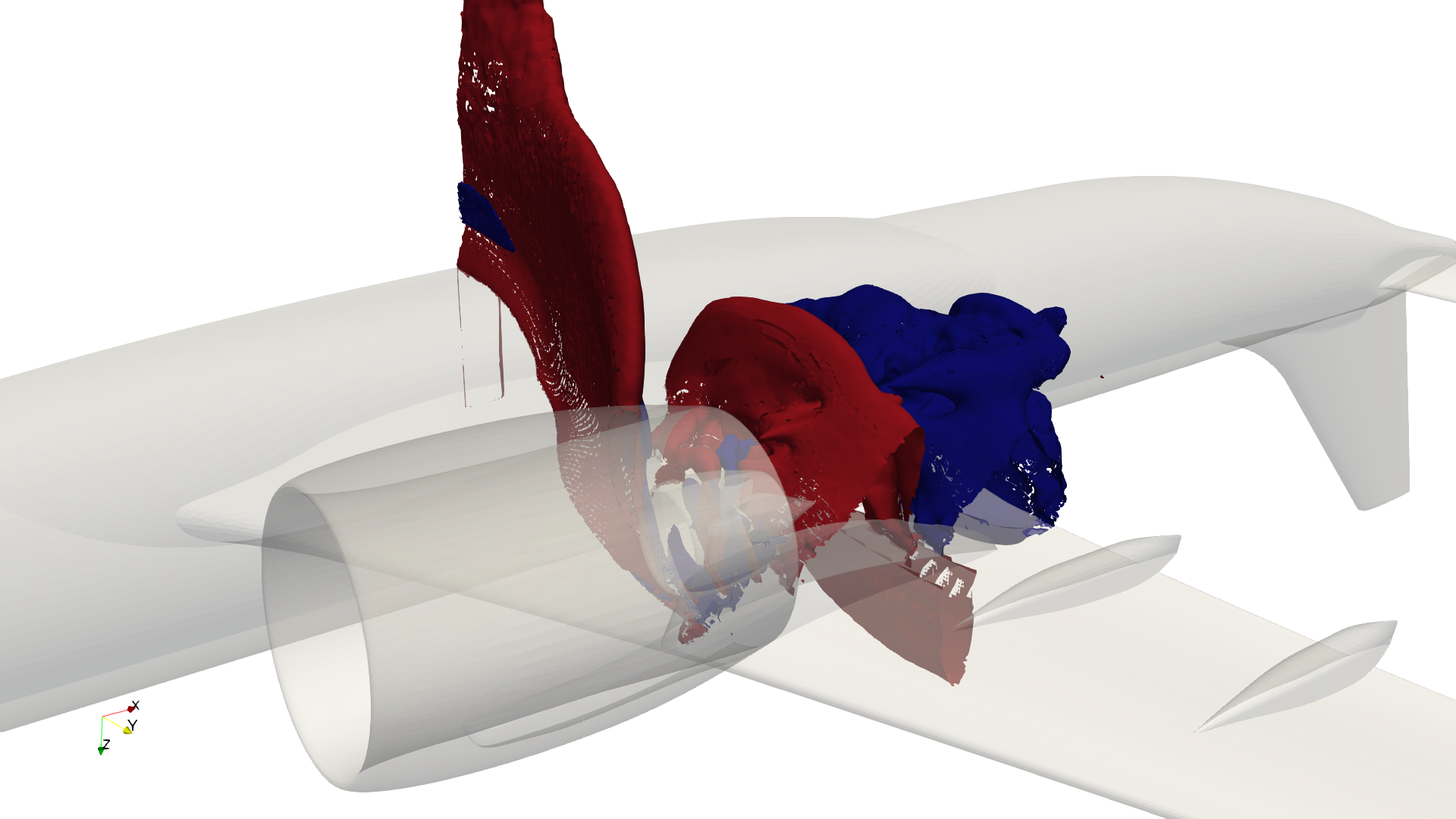}}
        \caption{$\tau=0.5$}
        \label{fig:sub6}
    \end{subfigure}

    \caption{Time series evolution of dominant volume mode in frequency band \textbf{2} over one half period ($\tau=t/T$ - normalized time, t - time, T - period); View from below the aircraft; Isosurfaces show reconstruction of the mode at $\Delta p / \lambda_{0} = \pm 10$ (positive values in red, negative values in blue)}
    \label{fig:spod_vol_iso10_timeseries}
\end{figure}

To complement the volume analysis, Fig. \ref{fig:spod_vol_iso10_timeseries} visualizes the reconstruction of the leading SPOD mode in frequency band \textbf{2} via isosurfaces of normalized pressure fluctuations. The sequence illustrates the three-dimensional temporal evolution of the coherent structures and spans half a period. Due to the anti-symmetry of the harmonic mode, the isosurfaces at $\Delta p / \lambda_{0} = \pm 10$ invert signs but retain their topology in the second half-period, meaning this sequence captures the complete spatial evolution of the mode. 

The snapshots reveal inboard-propagating buffet cells along the shock between pylon and fuselage on the lower surface, alongside downstream convection of shear layer structures. Upstream-traveling acoustic waves are identifiable on the upper surface (visible through the partially transparent geometry), confirming pressure wave propagation against the mean flow.
The phase progression confirms the traveling wave nature of the buffet phenomenon, demonstrating the spatial coupling between shock oscillations and wake dynamics. 
The visualizations were created employing the Sparse Spatial Sampling algorithm by Geise \cite{Geise2025}.

\section{Discussion}
\label{sec:discussion}
The surface and volumetric mode analyses provide a comprehensive understanding of the complex flow phenomena associated with the shock-induced separation on the wing lower surface. The leading modes identified in the frequency bands reveal wave-like shock motions, resembling buffet-cells, that originate at the pylon and propagate inboard towards the fuselage, correlated with pressure fluctuations on the surface in the rear part of the airfoil. This suggests a link between the shock oscillation and vortex shedding at the same frequency. Furthermore, the analysis of the phase angle gradients indicates that upstream traveling pressure waves are present both below and above the wing surface. Below the wing, these disturbances are deflected at the lower end of the shock and enter the supersonic region at an oblique angle, traveling from below toward the leading edge into the supersonic area, without penetrating the shock itself.

The volumetric mode analysis, particularly for the leading mode in frequency band \textbf{3}, highlights the significant role of shear layer breakup in generating pressure signals that propagate forward and away from the wall. This shear layer breakup is likely modulated by the pressure waves imposed by the shock oscillations, creating a complex interplay between the shock motion, shear layer instability, and vortex shedding. 

The observation of 2D-like shock oscillations for the mode in frequency band \textbf{2}, in areas where the shock is perpendicular to the mean flow, and the occurrence of buffet cells inboard where the shock is oblique, is consistent with the findings of Lusher \cite{Lusher2026}. They demonstrated that for low sweep angles, 2D-like shock oscillations dominate over 3D buffet-cells, while for larger sweep angles, 3D buffet-cells become more dominant. Transferred to our case, we interpret the angle between the shock and the mean flow as a local sweep angle in the sense of the study mentioned above.

The mode in frequency band \textbf{1} encodes shock oscillations that propagate inboard and outboard from a position that coincides with the spanwise boundary of the separation region near the fuselage. The separation region not only grows and shrinks ("breathes") in chordwise but also in spanwise direction. However, the spanwise oscillations exhibit a lower frequency.

The modal analysis reveals that all the identified phenomena occur simultaneously and overlap, influencing each other without a single causal factor being identified. The shock oscillation, buffet-cell propagation, vortex shedding, breakup of the shear layer, breathing of the separation region, and pressure wave propagation are all interconnected, creating a complex network of interactions that govern the flow behavior. The presence of multiple frequency bands, each corresponding to multiple flow phenomena, highlights the multifaceted nature of the shock-induced separation and the need for a comprehensive understanding of the underlying mechanisms.

A persistent challenge in our research has been the inability to resolve the shock buffet phenomenon using URANS approaches, despite extensive efforts to optimize modeling and numerical settings. This experience is consistent with the present findings, which suggest that capturing the unsteady vortex shedding and acoustic-pressure wave radiation downstream of the shock-induced separation is crucial for accurately reproducing the buffet phenomenon. The limitations of URANS in this regard are likely due to its tendency to produce large, quasi-steady separation bubbles that fail to generate coherent vortex shedding, a key feature of the shock buffet dynamics. In contrast, scale-resolving methods like DDES have been shown to be effective in capturing the complex, unsteady flow behavior associated with shock buffet, as they permit the explicit simulation of detached vortical structures in the wake of the shock-boundary-layer interaction and the resulting pressure-wave emission at the trailing edge. While it is possible that the inability to resolve shock buffet with URANS may be case-dependent and specific to our particular setup, the present results underscore the importance of using scale-resolving methods to accurately capture the underlying physics of this complex phenomenon.

\section{Conclusion}
In this study, we examined UHBR‑induced buffet on the wing lower surface by the combined analysis of high‑fidelity DDES and unsteady PSP measurements on a complex transport aircraft configuration. The DDES results exhibited excellent agreement with wind‑tunnel data and achieved an accuracy comparable to the considerably more elaborate IDDES simulations reported previously. Modal analysis based on the adaptive taper‑based SPOD identified the dominant flow structures associated with buffet frequencies in the Strouhal‑number range \(St\in[0.1,0.3]\).

Surface‑mode analysis revealed wave‑like shock motions that originate in the area of the pylon wing intersection and propagate inboard toward the fuselage. It was found that these shock motions correlate with pressure fluctuations on the rear portion of the airfoil, indicating a direct link between shock oscillation and vortex shedding at the same frequency. Modal analysis of the volumetric DDES pressure field supports this interpretation, showing a clear correlation between shock-induced pressure changes and downstream propagating structures in the separated flow region, while also identifying upstream traveling pressure waves on both the upper and lower wing surfaces. This is interpreted as a distinct nacelle-induced lower surface shock oscillation, yet similarities are observed with classical buffet on the wing upper surface at high incidences. It is therefore suggested that the underlying physical mechanism may be analogous, despite the different boundary conditions. 

The present investigation is confined to a Reynolds number of $Re=3.3\times10^{6}$, and the identified buffet mechanism must be validated at higher Reynolds numbers. Future work will therefore extend the combined DDES–SPOD framework to progressively larger \(Re\) regimes to assess the robustness and scalability of the proposed mechanism across realistic flight conditions.

To shed further light on the interplay among shock oscillations, flow separation, shear layer instabilities, and pressure waves, a global stability analysis of this configuration would be an interesting next step.
The stability analysis could show if any of the described mechanisms can appear in isolation, i.e., as a global, unstable mode.

\section*{Acknowledgement}
The authors gratefully acknowledge the Deutsche Forschungsgemeinschaft DFG (German Research Foundation) for funding this work in the research unit FOR 2895. The authors would like to thank the Helmholtz Gemeinschaft HGF (Helmholtz Association), Deutsches Zentrum für Luft- und Raumfahrt DLR (German Aerospace Center), and Airbus for providing the wind tunnel model and financing the wind tunnel measurements as well as public support to mature the test methods applied by DLR and ETW.

The authors gratefully acknowledge the scientific support and HPC resources provided by the German Aerospace Center (DLR). The HPC system CARA is partially funded by "Saxon State Ministry for Economic Affairs, Labour and Transport" and "Federal Ministry for Economic Affairs and Climate Action".

\section*{Declaration of generative AI and AI-assisted technologies in the manuscript preparation process}
During the preparation of this manuscript the authors employed the Chat‑AI Service of GWDG (Gesellschaft für wissenschaftliche Datenverarbeitung mbH Göttingen) to re‑phrase selected sentences and paragraphs for brevity and clarity and to improve the overall readability of the text. After using this service the authors carefully reviewed and edited the resulting material as needed and assume full responsibility for the content of the published article.

\printbibliography

\end{document}